\documentclass[twocolumn]{autart}    

\usepackage{graphicx} 
\usepackage{amsmath} 
\usepackage{amssymb}  
\usepackage{lipsum}
\usepackage{xcolor}
\usepackage{color}
\usepackage{cite}
\usepackage[normalem]{ulem}
\usepackage{xcolor,bbm, dsfont}
\usepackage{cite}
\usepackage{amsmath,amssymb,amsfonts}
\usepackage{graphicx}
\usepackage{graphicx}          

\renewcommand{\nuc}{\newcommand}
\nuc{\renuc}{\renewcommand}
\nuc{\dmo}{\DeclareMathOperator}

\nuc{\tcr}\textcopyright

\nuc{\noinsec}{\renewcommand{\theequation}{\arabic{section}.\arabic{equation}}}
\nuc{\noinapx}{\renewcommand{\theequation}{\Alph{section}.\arabic{equation}}}
\nuc{\noleminapx}{}
\nuc{\mysec}[1]{\section{#1}\setcounter{equation}{0}}

\nuc{\myapp}[1]{\section{#1}\setcounter{equation}{0}\noinapx\setcounter{lemma}{0}}

\newtheorem{alem}{Lemma}
\newtheorem{blem}{Lemma}
\newtheorem{clem}{Lemma}
\newtheorem{ass}{}

\nuc{\subsec}{\subsection}

\nuc{\ISd}{Itakura-Saito divergence}
\nuc{\pp}{point process}
\nuc{\bb}{Bhattacharyya bound}
\nuc{\cvd}{covariance density}
\nuc{\psd}{power spectral density}
\nuc{\BD}{Bregman divergence}
\nuc{\ifunc}{intensity function}
\nuc{\la}{local alternative}

\nuc{\Pp}{Point process}
\nuc{\Psp}{Poisson process}
\nuc{\rp}{renewal process}
\nuc{\Rnp}{Renewal process}
\nuc{\Hp}{Hawkes process}
\nuc{\HL}{Hawkes-Laguerre}
\nuc{\Lt}{Laplace transform}
\nuc{\iLt}{inverse Laplace transform}
\nuc{\iets}{interevent times}
\nuc{\Gd}{Gamma distributed}
\nuc{\lhf}{likelihood function}
\nuc{\lh}{likelihood}
\nuc{\tiv}{time-invariant}
\nuc{\tv}{time-varying}
\nuc{\Bhatt}{Bhattacharyya}
\nuc{\Bnl}{Bernoulli}
\nuc{\cpi}{counting process increment}

\nuc{\cme}{concentrated matrix exponential}
\nuc{\se}{self-exciting}
\nuc{\AWf}{Abate-Whitt framework}
\nuc{\mgf}{moment generating function}
\nuc{\KT}{Karamata Tauberian}
\nuc{\lyap}{Lyapunov }
\nuc{\iden}{identification }
\nuc{\sfr}{S$^4$}
\nuc{\sfv}{S$^5$}

\nuc{\rsi}{Riemann-Stieltjes integral}
\nuc{\RSi}{Riemann-Stieltjes integral}
\nuc{\RSc}{Riemann-Stieltjes convolution}
\nuc{\CSi}{Cauchy-Schwarz inequality}

\nuc{\ul}{\underline}

\nuc{\ra}{\rightarrow}
\nuc{\Ra}{\Rightarrow}
\nuc{\RA}{\Longrightarrow}
\nuc{\La}{\Leftarrow}
\nuc{\xra}[1]{\xrightarrow{#1}}
\nuc{\LRa}{\Leftrightarrow}
\nuc{\toi}{\to\infty}
\nuc{\trid}{\triangledown}
\nuc{\triu}{\triangleup}
\nuc{\bs}{\backslash}
\nuc{\convas}{\xra{a.s.}}
\nuc{\convp}{\xra{p}}
\nuc{\convd}{\xra{d}}
\nuc{\trieq}{\triangleq}

\nuc{\twn}{t_1^n}
\nuc{\xwn}{x_1^n}
\nuc{\Snoi}{\sum_{n=0}^\infty}
\nuc{\Snwi}{\sum_{n=1}^\infty}
\nuc{\swn}{\ssum1^n}
\nuc{\swN}{\ssum1^N}
\nuc{\swP}{\ssum1^P}
\nuc{\soN}{\ssum0^N}
\nuc{\swm}{\ssum1^m}
\nuc{\swM}{\ssum1^M}
\nuc{\siwn}{\ssum{i=1}^n}
\nuc{\siwnl}{\ssum{i=1}^{n_l}}
\nuc{\srwn}{\ssum{r=1}^n}
\nuc{\srwnl}{\ssum{r=1}^{n_l}}
\nuc{\srwnm}{\ssum{r=1}^{n_m}}
\nuc{\sawM}{\ssum{a=1}^M}
\nuc{\sbwM}{\ssum{b=1}^M}
\nuc{\siwm}{\ssum{i=1}^m}
\nuc{\skwm}{\ssum{k=1}^m}
\nuc{\siwM}{\ssum{i=1}^M}
\nuc{\skwM}{\ssum{k=1}^M}
\nuc{\siwK}{\ssum{i=1}^K}
\nuc{\sjwm}{\ssum{j=1}^m}
\nuc{\sjwM}{\ssum{j=1}^M}
\nuc{\slwM}{\ssum{l=1}^M}
\nuc{\smwM}{\ssum{m=1}^M}
\nuc{\sjwp}{\ssum{j=1}^p}
\nuc{\sjwP}{\ssum{j=1}^P}
\nuc{\swk}{\ssum1^k}
\nuc{\soi}{\ssum0^\infty}
\nuc{\swi}{\ssum1^\infty}
\nuc{\pwn}{\Pi_1^n}
\nuc{\pwk}{\Pi_1^k}
\nuc{\pwm}{\Pi_1^m}
\nuc{\prwn}{\Pi_{r=1}^n}
\nuc{\TwN}{T_1^N}

\nuc{\Stiltr}{\sum_{\ti<\tr}}
\nuc{\Sticlltrcm}{\sum_{\ticl<\trcm}}

\nuc{\Swn}{\sum_1^n}
\nuc{\Swm}{\sum_1^m}
\nuc{\SwM}{\sum_1^M}
\nuc{\Swk}{\sum_1^k}
\nuc{\Pwn}{\prod_1^n}
\nuc{\Pwm}{\prod_1^m}
\nuc{\Swp}{\sum_1^p}
\nuc{\SwP}{\sum_1^P}

\nuc{\intpi}{\int_{-\pi}^{\pi}}
\nuc{\intii}{\int_{-\infty}^{\infty}}
\nuc{\intoT}{\int_0^T}
\nuc{\intox}{\int_0^x}
\nuc{\intxi}{\int_x^\infty}
\nuc{\intti}{\int_t^\infty}
\nuc{\intit}{\int_{-\infty}^t}
\nuc{\intio}{\int_{-\infty}^0}
\nuc{\intiT}{\int_{-\infty}^T}
\nuc{\intot}{\int_0^t}
\nuc{\intou}{\int_0^u}
\nuc{\intoinfty}{\int_0^\infty}
\nuc{\intoi}{\int_0^\infty}
\nuc{\intmp}{\intinfty}

\nuc{\limto}{\lim_{t\to0}}
\nuc{\limTo}{\lim_{T\to0}}
\nuc{\limxo}{\lim_{x\to0}}
\nuc{\limso}{\lim_{s\to0}}
\nuc{\stoo}{s\to0}
\nuc{\stoi}{s\to\infty}
\nuc{\xtoi}{x\to\infty}
\nuc{\xtoo}{x\to0}
\nuc{\Ttoi}{T\to\infty}
\nuc{\ntoi}{n\to\infty}

\nuc{\limti}{\lim_{t\to\infty}}
\nuc{\limTi}{\lim_{T\to\infty}}
\nuc{\limxi}{\lim_{x\to\infty}}
\nuc{\limsi}{\lim_{s\to\infty}}

\nuc{\st}{\mbox{s.t. }}

\nuc{\hlamt}{\hat{\lambda}(t)}

\nuc{\nid}{n_i^\delta}
\nuc{\Nid}{N_i^\delta}

\nuc{\intsum}{\ssum0^\infty\int_{R_n(T)}}
\nuc{\ints}{\ssum0^\infty\int}

\nuc{\lt}{\left}
\nuc{\rt}{\right}

\nuc{\wos}{\frac1s}

\nuc{\wot}{\frac1t}
\nuc{\woT}{\frac1T}
\nuc{\wo}[1]{\frac1{#1}}
\nuc{\wox}{\wo}
\nuc{\haf}{{\frac12}}
\nuc{\mhaf}{{-\frac12}}

\nuc{\cH}{\mathcal{H}}
\nuc{\cB}{\mathcal{B}}
\nuc{\cD}{\mathcal{D}}
\nuc{\cN}{\mathcal{N}}
\nuc{\cP}{\mathcal{P}}
\nuc{\calj}{{\cal J}}
\nuc{\cHT}{\cH_T}
\nuc{\cO}{\mathcal{O}}
\nuc{\cL}{\mathcal{L}}
\nuc{\bR}{\mathbb R}
\nuc{\bC}{\mathbb C}
\nuc{\cl}{\mathcal{l}}
\nuc{\cI}{\mathcal{I}}
\nuc{\bbR}{\mathbb{R}}
\nuc{\bN}{\mathbb{N}}
\nuc{\bZ}{\mathbb{Z}}
\nuc{\bbw}[1]{\mathbbm{1}_{#1}}
\nuc{\N}{\mathfrak{N}}
\nuc{\n}{\mathfrak{n}}
\nuc{\bx}{{\bf x}}
\nuc{\by}{{\bf y}}
\nuc{\subs}{\subset}
\nuc{\wnb}{{\bf 1}}

\nuc{\eq}[1]{\begin{align*}#1\end{align*}}
\nuc{\eqn}[1]{\begin{align}#1\end{align}}
\nuc{\bmat}[1]{\begin{bmatrix}#1\end{bmatrix}}
\nuc{\iary}[2]{\begin{IEEEeqnarray*}{#1}#2\end{IEEEeqnarray*}}
\nuc{\iaryn}[2]{\begin{IEEEeqnarray}{#1}#2\end{IEEEeqnarray}}
\nuc{\subeq}[1]{\begin{subequations}#1\end{subequations}}
\nuc{\smat}[1]{\begin{smallmatrix}#1\end{smallmatrix}}
\nuc{\mat}[1]{\begin{matrix}#1\end{matrix}}
\nuc{\sqmat}[1]{\sqbra{\begin{matrix}#1\end{matrix}}}
\nuc{\chuz}[2]{\bra{\mat{#1\\#2}}}

\newcommand{\BNu}{\begin{enumerate}}
\newcommand{\ENu}{\end{enumerate}}
\newcommand{\Bit}{\begin{itemize}}
\newcommand{\Eit}{\end{itemize}}
\newcommand{\Bres}[1]{\begin{result}\label{#1}}
\newcommand{\Eres}{\end{result}}
\newcommand{\Bdef}[1]{\begin{defn}\label{#1}}
\newcommand{\Edef}{\end{defn}}
\newcommand{\Blem}[1]{\begin{lem}\label{#1}}
\newcommand{\Elem}{\end{lem}}
\newcommand{\Balem}[1]{\begin{alem}\label{#1}}
\newcommand{\Ealem}{\end{alem}}
\newcommand{\Bblem}[1]{\begin{blem}\label{#1}}
\newcommand{\Eblem}{\end{blem}}
\newcommand{\Bclem}[1]{\begin{clem}\label{#1}}
\newcommand{\Eclem}{\end{clem}}
\newcommand{\Brem}[1]{\begin{rem}\label{#1}}
\newcommand{\Erem}{\end{rem}}
\newcommand{\Bthm}[1]{\begin{thm}\label{#1}}
\newcommand{\Ethm}{\end{thm}}
\newcommand{\Bass}[1]{\begin{ass}\label{#1}}
\newcommand{\Eass}{\end{ass}}
\newcommand{\Balg}[1]{\begin{alg}\label{#1}}
\newcommand{\Ealg}{\end{alg}}
\newcommand{\Bprop}[1]{\begin{prop}\label{#1}}
\newcommand{\Eprop}{\end{prop}}
\newcommand{\Bcor}[1]{\begin{coro}\label{#1}}
\newcommand{\Ecor}{\end{coro}}
\newcommand{\Bpf}{\begin{pf*}}
\newcommand{\Epf}{{\hfill$\square$}\end{pf*}}
\newcommand{\Bsmat}[1]{\begin{smallmatrix}#1}
\newcommand{\Esmat}{\end{smallmatrix}}
\newcommand{\Bmat}[1]{\begin{matrix}#1}
\newcommand{\Emat}{\end{matrix}}
\nuc{\Brmk}{\begin{remark}}
\nuc{\Brmks}{\begin{remarks}}
\nuc{\Ermk}{\end{remark}}
\nuc{\Ermks}{\end{remarks}}

\nuc{\emu}[1]{e^{-#1}}

\nuc{\theo}[1]{\begin{thm}#1\end{thm}}
\nuc{\defi}[1]{\begin{defn}#1\end{defn}}
\nuc{\pro}[1]{\begin{proof}#1\end{proof}}
\nuc{\cas}[1]{\begin{cases}#1\end{cases}}
\nuc{\arr}[2]{\begin{array}{#1}#2\end{array}}
\nuc{\bra}[1]{\left(#1\right)}
\nuc{\sqbra}[1]{\left[#1\right]}
\nuc{\ang}[1]{\langle#1\rangle}
\nuc{\floor}[1]{\lfloor#1\rfloor}
\nuc{\Ver}[1]{\lVert#1\rVert}
\nuc{\ver}[1]{\lvert#1\rvert}
\nuc{\Bver}[1]{\Bigl\vert#1\Bigr\vert}
\nuc{\bver}[1]{\bigl\vert#1\bigr\vert}
\nuc{\BVer}[1]{\Bigl\Vert#1\Bigr\Vert}
\nuc{\bVer}[1]{\bigl\Vert#1\bigr\Vert}
\nuc{\ssum}[1]{{\textstyle\sum}_{#1}}
\nuc{\sprod}[1]{{\textstyle\prod}_{#1}}
\nuc{\inv}[1]{#1^{-1}}
\nuc{\ddt}[1]{\frac{d#1}{dt}}
\nuc{\ddx}[2]{\frac{d#1}{d#2}}
\nuc{\ppt}[1]{\frac{\partial#1}{\partial t}}
\nuc{\ppx}[2]{\frac{\partial#1}{\partial#2}}

\nuc{\sbij}[1]{#1_{ij}}
\nuc{\sbbij}[1]{\bar #1_{ij}}
\nuc{\sbi}[1]{#1_{i}}
\nuc{\sbbi}[1]{\bar #1_{i}}
\nuc{\sbj}[1]{#1_{j}}
\nuc{\sbbj}[1]{\bar #1_{j}}
\nuc{\nfd}[1]{#1^{(n)}}
\nuc{\limo}[1]{\lim_{#1\to0}}
\nuc{\limi}[1]{\lim_{#1\to\infty}}

\nuc{\grad}{\bigtriangledown}
\nuc{\gradx}[1]{\bigtriangledown_{#1}}
\nuc{\hessx}[1]{\bigtriangledown^2_{#1}}

\nuc{\idc}[1]{\mathbbm{1}_{#1}}

\newcommand{\nlist}[1]{\begin{enumerate}#1\end{enumerate}}

\nuc{\itum}[1]{\item[(#1)]}
\nuc{\ita}{\itum{a}}
\nuc{\itb}{\itum{b}}
\nuc{\itc}{\itum{c}}
\nuc{\itd}{\itum{d}}
\nuc{\ite}{\itum{e}}
\nuc{\itf}{\itum{f}}
\nuc{\itg}{\itum{g}}
\nuc{\ith}{\itum{h}}

\DeclareMathOperator{\E}{E}
\DeclareMathOperator{\Var}{Var}

\DeclareMathOperator{\Cov}{Cov}

\DeclareMathOperator*{\argmin}{arg\,min}

\nuc{\ai}{a_i}
\nuc{\ak}{a_k}
\nuc{\aj}{a_j}
\nuc{\aij}{a_{ij}}
\nuc{\akl}{a_{kl}}
\nuc{\akj}{a_{kj}}
\nuc{\abi}{\bar a_i}
\nuc{\abk}{\bar a_k}
\nuc{\abj}{\bar a_j}
\nuc{\abij}{\bar a_{ij}}
\nuc{\abkj}{\bar a_{kj}}
\nuc{\alpm}{\alpha_{m}}
\nuc{\alpi}{\alpha_{i}}
\nuc{\alpk}{\alpha_{k}}
\nuc{\alpj}{\alpha_{j}}
\nuc{\alpij}{\alpha_{ij}}
\nuc{\alpkj}{\alpha_{kj}}
\nuc{\alpab}{\alpha_{ab}}
\nuc{\Ahw}{\hat A_1}
\nuc{\Aho}{\hat A_0}
\nuc{\ajcl}{a_{j,l}}
\nuc{\ajclcm}{a_{j,l,m}}
\nuc{\Ahat}{\hat A}
\nuc{\Ab}{A_b}

\nuc{\Ah}{\hat{A}}
\nuc{\alp}{\alpha}

\nuc{\Bi}{B_i}
\nuc{\Bj}{B_j}
\nuc{\Bij}{B_{ij}}
\nuc{\Bkj}{B_{kj}}
\nuc{\bij}{b_{ij}}
\nuc{\bkj}{b_{kj}}
\nuc{\Bbi}{\bar B_i}
\nuc{\Bbj}{\bar B_j}
\nuc{\Bbij}{\bar B_{ij}}
\nuc{\Bbkj}{\bar B_{kj}}
\nuc{\beti}{\beta_i}
\nuc{\betj}{\beta_j}
\nuc{\betk}{\beta_k}
\nuc{\betaa}{\beta_a}
\nuc{\betb}{\beta_b}
\nuc{\betij}{\beta_{ij}}
\nuc{\betkl}{\beta_{kl}}
\nuc{\betkj}{\beta_{kj}}
\nuc{\betab}{\beta_{ab}}
\nuc{\betjcl}{\beta_{j,l}}
\nuc{\betjclcm}{\beta_{j,l,m}}
\nuc{\Bhw}{\hat B_1}
\nuc{\Bho}{\hat B_0}

\nuc{\ci}{c_i}
\nuc{\cj}{c_j}
\nuc{\ck}{c_k}
\nuc{\ca}{c_a}
\nuc{\cb}{c_b}

\nuc{\cm}{c_m}
\nuc{\cij}{c_{ij}}
\nuc{\ckj}{c_{kj}}
\nuc{\cab}{c_{ab}}
\nuc{\cbi}{\bar c_i}
\nuc{\cbj}{\bar c_j}
\nuc{\Ci}{C_i}
\nuc{\Cj}{C_j}
\nuc{\Cij}{C_{ij}}
\nuc{\Cbi}{\bar C_i}
\nuc{\Cbj}{\bar C_j}
\nuc{\Chw}{\hat C_1}
\nuc{\Cho}{\hat C_0}
\nuc{\cwt}{c_{12}}
\nuc{\Chat}{\hat C}

\nuc{\Di}{D_i}
\nuc{\Dj}{D_j}
\nuc{\Dbi}{\bar D_i}
\nuc{\Dbj}{\bar D_j}
\nuc{\Dij}{D_{ij}}
\nuc{\Dkj}{D_{kj}}
\nuc{\Dab}{D_{ab}}
\nuc{\dij}{d_{ij}}
\nuc{\dab}{d_{ab}}
\nuc{\dkj}{d_{kj}}
\nuc{\Dbij}{\bar D_{ij}}
\nuc{\Dbkj}{\bar D_{kj}}
\nuc{\Dbab}{\bar D_{ab}}
\nuc{\DT}{D_T}

\nuc{\et}{e_t}
\nuc{\ep}{\epsilon}

\nuc{\fk}{f_k}
\nuc{\Fi}{F_i}
\nuc{\Fk}{F_k}
\nuc{\Fj}{F_j}
\nuc{\Fa}{F_a}
\nuc{\Fb}{F_b}
\nuc{\Fm}{F_m}
\nuc{\Ft}{F_t}
\nuc{\Fbi}{\bar F_i}
\nuc{\Fbj}{\bar F_j}
\nuc{\Fba}{\bar F_a}
\nuc{\Fbb}{\bar F_b}
\nuc{\Fij}{F_{ij}}
\nuc{\Fkj}{F_{kj}}
\nuc{\Fab}{F_{ab}}
\nuc{\fij}{f_{ij}}
\nuc{\fkj}{f_{kj}}
\nuc{\fab}{f_{ab}}
\nuc{\Fbij}{\bar F_{ij}}
\nuc{\Fbkj}{\bar F_{kj}}
\nuc{\Fbab}{\bar F_{ab}}
\nuc{\Fwtu}{F_{12}}
\nuc{\Fbwtu}{\bar F_{12}}
\nuc{\Fpct}{F_{p,t}}
\nuc{\Fh}{\hat F}
\nuc{\Fhat}{\hat F}

\nuc{\gam}{\gamma}
\nuc{\Gam}{\Gamma}
\nuc{\gami}{\gam_i}
\nuc{\gamk}{\gam_k}
\nuc{\gamj}{\gam_j}
\nuc{\gamij}{\gam_{ij}}
\nuc{\gamkj}{\gam_{kj}}
\nuc{\gamab}{\gam_{ab}}
\nuc{\Gi}{G_i}
\nuc{\Gk}{G_k}
\nuc{\Gj}{G_j}
\nuc{\Ga}{G_a}
\nuc{\Gb}{G_b}
\nuc{\Gbi}{\bar G_i}
\nuc{\Gbj}{\bar G_j}
\nuc{\Gij}{G_{ij}}
\nuc{\Gkj}{G_{kj}}
\nuc{\Gab}{G_{ab}}
\nuc{\Gbab}{\bar G_{ab}}
\nuc{\Gbij}{\bar G_{ij}}
\nuc{\Gbkj}{\bar G_{kj}}
\nuc{\Gwtu}{G_{12}}
\nuc{\Gbwtu}{\bar G_{12}}
\nuc{\bGwt}{\bar G_{12}}
\nuc{\Gwt}{G_{12}}

\nuc{\bgwt}{\bar g_{12}}

\nuc{\gwt}{g_{12}}

\nuc{\hi}{h_i}
\nuc{\hk}{h_k}
\nuc{\hj}{h_j}
\nuc{\ha}{h_a}
\nuc{\hb}{h_b}
\nuc{\hd}{h_d}
\nuc{\hbi}{\bar h_i}
\nuc{\hbk}{\bar h_k}
\nuc{\hbj}{\bar h_j}
\nuc{\hba}{\bar h_a}
\nuc{\hbb}{\bar h_b}
\nuc{\hij}{h_{ij}}
\nuc{\hbij}{\bar h_{ij}}
\nuc{\hkj}{h_{kj}}
\nuc{\hbkj}{\bar h_{kj}}
\nuc{\hab}{h_{ab}}
\nuc{\hbab}{\bar h_{ab}}
\nuc{\hjcm}{h_{j,m}}
\nuc{\hjcl}{h_{j,l}}
\nuc{\hjclcm}{h_{j,l,m}}
\nuc{\Hi}{H_i}
\nuc{\Hk}{H_k}
\nuc{\Hj}{H_j}
\nuc{\Hbi}{\bar H_i}
\nuc{\Hbk}{\bar H_k}
\nuc{\Hbj}{\bar H_j}
\nuc{\Hij}{H_{ij}}
\nuc{\Hbij}{\bar H_{ij}}
\nuc{\Hkj}{H_{kj}}
\nuc{\Hbkj}{\bar H_{kj}}
\nuc{\Hab}{H_{kj}}
\nuc{\Hbab}{\bar H_{ab}}
\nuc{\Hwtu}{H_{12}}
\nuc{\Hbwtu}{\bar H_{12}}
\nuc{\Hinf}{\cH_\infty}
\nuc{\cHTicj}{\cH_{T,i,j}}
\nuc{\calhr}{{\cal H}_R}

\nuc{\jw}{\jmath\omega}

\nuc{\kapij}{\kappa_{ij}}
\nuc{\kapbij}{\bar\kappa_{ij}}
\nuc{\kapkj}{\kappa_{kj}}
\nuc{\kapbkj}{\bar\kappa_{kj}}
\nuc{\kapab}{\kappa_{ab}}
\nuc{\kapbab}{\bar\kappa_{ab}}
\nuc{\kapji}{\kappa_{ji}}
\nuc{\kapbji}{\bar\kappa_{ji}}
\nuc{\Ki}{K_i}
\nuc{\Kk}{K_k}
\nuc{\Kj}{K_j}
\nuc{\Kt}{K_t}
\nuc{\Kbi}{\bar K_i}
\nuc{\Kbk}{\bar K_k}
\nuc{\Kbj}{\bar K_j}
\nuc{\Kij}{K_{ij}}
\nuc{\Kbij}{\bar K_{ij}}
\nuc{\Kkj}{K_{kj}}
\nuc{\Kbkj}{\bar K_{kj}}
\nuc{\Kab}{K_{ab}}
\nuc{\Kbab}{\bar K_{ab}}
\nuc{\Kji}{K_{ji}}
\nuc{\Kbji}{\bar K_{ji}}
\nuc{\Kwtu}{K_{12}}
\nuc{\Kbwtu}{\bar K_{12}}
\nuc{\calhk}{{\cal H}_K}

\nuc{\lam}{\lambda}
\nuc{\Lam}{\Lambda}
\nuc{\lami}{\lambda_{i}}
\nuc{\lamk}{\lambda_{k}}
\nuc{\lamj}{\lambda_{j}}
\nuc{\lama}{\lambda_{a}}
\nuc{\lamb}{\lambda_{b}}
\nuc{\lamm}{\lambda_{m}}
\nuc{\laml}{\lambda_{l}}
\nuc{\lamij}{\lambda_{ij}}
\nuc{\lamkj}{\lambda_{kj}}
\nuc{\lamab}{\lambda_{ab}}
\nuc{\lamicj}{\lambda_{i,j}}
\nuc{\lambij}{\bar\lambda_{ij}}
\nuc{\lambab}{\bar\lambda_{ab}}
\nuc{\Lami}{\Lambda_{i}}
\nuc{\Lamk}{\Lambda_{k}}
\nuc{\Lama}{\Lambda_{a}}
\nuc{\Lamj}{\Lambda_{j}}
\nuc{\Lamij}{\Lambda_{ij}}
\nuc{\Lamkj}{\Lambda_{kj}}
\nuc{\Lamab}{\Lambda_{ab}}
\nuc{\Lamicj}{\Lambda_{i,j}}
\nuc{\Lambij}{\bar\Lambda_{ij}}

\nuc{\lamTt}{\lam_t^T}
\nuc{\lamtgT}{\lam_{t|T}}
\nuc{\lamTu}{\lam_u^T}
\nuc{\lamugT}{\lam_{u|T}}
\nuc{\lamicjclcm}{\lam_{i,j,l,m}}
\nuc{\lamTtcu}{\lam_{t,u}^T}
\nuc{\lamtcugT}{\lam_{t,u|T}}
\nuc{\lamTtpw}{\lam^T_{t+1}}
\nuc{\lamtpwgT}{\lam_{t+1|T}}
\nuc{\LamTt}{\Lam_t^T}
\nuc{\LamtgT}{\Lam_{t|T}}
\nuc{\LamTu}{\Lam_u^T}
\nuc{\LamugT}{\Lam_{u|T}}
\nuc{\LamTtpw}{\Lam^T_{t+1}}
\nuc{\LamtpwgT}{\Lam_{t+1|T}}
\nuc{\LamTtcu}{\Lam_{t,u}^T}
\nuc{\LamtcugT}{\Lam_{t,u|T}}

\nuc{\Li}{L_i}
\nuc{\Lk}{L_k}
\nuc{\Lj}{L_j}
\nuc{\cLa}{\cL_a}
\nuc{\cLb}{\cL_b}

\nuc{\Lb}{L_b}
\nuc{\Lbi}{\bar L_i}
\nuc{\Lbk}{\bar L_k}
\nuc{\Lbj}{\bar L_j}
\nuc{\Lij}{L_{ij}}
\nuc{\Lkj}{L_{kj}}
\nuc{\Lab}{L_{ab}}
\nuc{\Licj}{L_{i,j}}
\nuc{\Lbij}{\bar L_{ij}}
\nuc{\Lbkj}{\bar L_{kj}}
\nuc{\Lwtu}{L_{12}}
\nuc{\Lbwtu}{\bar L_{12}}
\nuc{\LTtn}{L(T,n;t_1^n)}
\nuc{\mi}{m_i}
\nuc{\mk}{m_k}
\nuc{\mj}{m_j}
\nuc{\mij}{m_{ij}}
\nuc{\mkj}{m_{kj}}
\nuc{\mab}{m_{ab}}
\nuc{\Mij}{M_{ij}}
\nuc{\Mkj}{M_{kj}}
\nuc{\Mab}{M_{ab}}
\nuc{\Mfij}{M_{\fij}}
\nuc{\mui}{\mu_i}
\nuc{\muk}{\mu_k}
\nuc{\muj}{\mu_j}
\nuc{\mua}{\mu_a}
\nuc{\mub}{\mu_b}
\nuc{\muij}{\mu_{ij}}
\nuc{\mukj}{\mu_{kj}}
\nuc{\muab}{\mu_{ab}}
\nuc{\mbi}{\bar m_i}
\nuc{\mbk}{\bar m_k}
\nuc{\mbj}{\bar m_j}
\nuc{\mbij}{\bar m_{ij}}
\nuc{\mbkj}{\bar m_{kj}}

\nuc{\NoT}{N_0^T}
\nuc{\nut}{\nu_t}
\nuc{\nl}{n_l}
\nuc{\nm}{n_m}
\nuc{\Nt}{N_t}
\nuc{\NT}{N_T}
\nuc{\nT}{n_T}
\nuc{\nicj}{n_{i,j}}
\nuc{\Nu}{N_u}
\nuc{\Nucm}{N_{u,m}}
\nuc{\Nucl}{N_{u,l}}
\nuc{\Ntcl}{N_{t,l}}
\nuc{\Not}{N_0^t}
\nuc{\Notm}{N_0^{t_-}}

\nuc{\ome}{\omega}
\nuc{\Ome}{\Omega}
\nuc{\omekl}{\ome_{kl}}
\nuc{\omeal}{\ome_{al}}
\nuc{\omet}{\ome_t}
\nuc{\Omeicj}{\Ome_{i,j}}
\nuc{\Omebicj}{\bar\Ome_{i,j}}

\nuc{\pbi}{\bar p_i}
\nuc{\pbk}{\bar p_k}
\nuc{\pbj}{\bar p_j}
\nuc{\pba}{\bar p_a}
\nuc{\pbb}{\bar p_b}
\nuc{\pik}{\pi_k}
\nuc{\pj}{p_j}
\nuc{\pk}{p_k}
\nuc{\pa}{p_a}
\nuc{\pb}{p_b}
\nuc{\pij}{p_{ij}}
\nuc{\pkj}{p_{kj}}
\nuc{\pab}{p_{ab}}
\nuc{\pbar}{\bar p}
\nuc{\pbij}{\bar p_{ij}}
\nuc{\pbkj}{\bar p_{kj}}
\nuc{\pbab}{\bar p_{ab}}
\nuc{\ptij}{\tilde p_{ij}}
\nuc{\pwtu}{p_{12}}
\nuc{\pbwtu}{\bar p_{12}}

\nuc{\Picj}{P_{i,j}}
\nuc{\PTicj}{P^T_{i,j}}
\nuc{\PTicipw}{P^T_{i,i+w}}
\nuc{\PTicipk}{P^T_{i,i+k}}

\nuc{\Pt}{P_t}
\nuc{\Pu}{P_u}
\nuc{\Ptcu}{P_{t,u}}
\nuc{\Ptck}{P_{t,k}}
\nuc{\PTtcu}{P^T_{t,u}}
\nuc{\PtcugT}{P_{t,u|T}}
\nuc{\Pstcu}{P^s_{t,u}}
\nuc{\Ptcugs}{P_{t,u|s}}
\nuc{\PTtctpw}{P^T_{t,t+w}}
\nuc{\PTtctpk}{P^T_{t,t+k}}
\nuc{\Ptt}{P^t_t}
\nuc{\Ptgt}{P_{t|t}}
\nuc{\PtgT}{P_{t|T}}
\nuc{\Pst}{P^s_t}
\nuc{\Ptgs}{P_{t|s}}
\nuc{\PTt}{P^T_t}
\nuc{\Ptmwt}{P^{t-1}_t}
\nuc{\Ptgtmw}{P_{t|t-1}}
\nuc{\Po}{P_0}
\nuc{\Poi}{P_0^{-1}}

\nuc{\phiad}{\phi_{ad}}
\nuc{\phiba}{\phi_{ba}}
\nuc{\phiaa}{\phi_{aa}}
\nuc{\phibd}{\phi_{bd}}
\nuc{\phicd}{\phi_{cd}}
\nuc{\phiadh}{\hat\phi_{ad}}
\nuc{\phibah}{\hat\phi_{ba}}
\nuc{\phiaah}{\hat\phi_{aa}}
\nuc{\phibdh}{\hat\phi_{bd}}
\nuc{\phicdh}{\hat\phi_{cd}}

\nuc{\bpwt}{\bar p_{12}}
\nuc{\pwt}{p_{12}}

\nuc{\Qhat}{\hat Q}
\nuc{\Qbi}{\bar Q_i}
\nuc{\Qbk}{\bar Q_k}
\nuc{\Qbj}{\bar Q_j}
\nuc{\Qj}{Q_j}
\nuc{\Qij}{Q_{ij}}
\nuc{\Qkj}{Q_{kj}}
\nuc{\Qab}{Q_{ab}}
\nuc{\Qm}{Q_m}
\nuc{\Qjclcm}{Q_{j,l,m}}
\nuc{\Qbij}{\bar Q_{ij}}
\nuc{\Qwtu}{Q_{12}}
\nuc{\Qbwtu}{\bar Q_{12}}
\nuc{\qbi}{\bar q_i}
\nuc{\qbk}{\bar q_k}
\nuc{\qbj}{\bar q_j}
\nuc{\qba}{\bar q_a}
\nuc{\qbb}{\bar q_b}
\nuc{\qj}{q_j}
\nuc{\qij}{q_{ij}}
\nuc{\qbij}{\bar q_{ij}}
\nuc{\qkj}{q_{kj}}
\nuc{\qbkj}{\bar q_{kj}}
\nuc{\qab}{q_{ab}}
\nuc{\qbab}{\bar q_{ab}}
\nuc{\qwtu}{q_{12}}
\nuc{\qbwtu}{\bar q_{12}}
\nuc{\phiT}{\phi_T}
\nuc{\bqwt}{\bar q_{12}}

\nuc{\qwt}{q_{12}}

\nuc{\Qv}{Q_v}
\nuc{\Qep}{Q_\ep}
\nuc{\Qw}{Q_w}

\nuc{\Rhat}{\hat R}
\nuc{\Rbi}{\bar R_i}
\nuc{\Rbj}{\bar R_j}
\nuc{\Ri}{R_i}
\nuc{\Rj}{R_j}
\nuc{\Rt}{R_t}
\nuc{\Rij}{R_{ij}}
\nuc{\Rkj}{R_{kj}}
\nuc{\Rab}{R_{ab}}
\nuc{\Rjclcm}{R_{j,l,m}}
\nuc{\rij}{r_{ij}}
\nuc{\Rbij}{\bar R_{ij}}
\nuc{\Rbkj}{\bar R_{kj}}
\nuc{\Rbab}{\bar R_{ab}}
\nuc{\Rwtu}{R_{12}}
\nuc{\Rbwtu}{\bar R_{12}}
\nuc{\rhoT}{\rho_T}
\nuc{\rhoi}{\rho_{i}}
\nuc{\rhok}{\rho_{k}}
\nuc{\rhoj}{\rho_{j}}
\nuc{\rhoa}{\rho_{a}}
\nuc{\rhob}{\rho_{b}}
\nuc{\rhoij}{\rho_{ij}}
\nuc{\rhokj}{\rho_{kj}}
\nuc{\rhoab}{\rho_{ab}}
\nuc{\Rect}{R_{e,t}}

\nuc{\Sig}{\Sigma}
\nuc{\sig}{\sigma}
\nuc{\Si}{S_i}
\nuc{\Sk}{S_k}
\nuc{\Sj}{S_j}
\nuc{\Sa}{S_a}
\nuc{\Sb}{S_b}
\nuc{\Sbar}{\bar S}
\nuc{\Sbi}{\bar S_i}
\nuc{\Sbk}{\bar S_k}
\nuc{\Sbj}{\bar S_j}
\nuc{\Sba}{\bar S_a}
\nuc{\Sbb}{\bar S_b}
\nuc{\Sij}{S_{ij}}
\nuc{\Sbij}{\bar S_{ij}}
\nuc{\Skj}{S_{kj}}
\nuc{\Sbkj}{\bar S_{kj}}
\nuc{\Sab}{S_{ab}}
\nuc{\Sbab}{\bar S_{ab}}
\nuc{\Swtu}{S_{12}}
\nuc{\Sww}{S_{11}}
\nuc{\Swo}{S_{10}}
\nuc{\Sow}{S_{01}}
\nuc{\Soo}{S_{00}}
\nuc{\Sbwtu}{\bar S_{12}}
\nuc{\Sigxx}{\ssum{xx}}
\nuc{\Sigx}{\ssum{x}}
\nuc{\Sigy}{\ssum{y}}
\nuc{\Sigxy}{\ssum{xy}}
\nuc{\Sigyx}{\ssum{yx}}
\nuc{\Sigyy}{\ssum{yy}}
\nuc{\Sxx}{S_{xx}}
\nuc{\Sx}{S_{x}}
\nuc{\Sy}{S_{x}}
\nuc{\Sxy}{S_{xy}}
\nuc{\Syx}{S_{yx}}
\nuc{\Syy}{S_{yy}}
\nuc{\Sjclcm}{S_{j,l,m}}
\nuc{\sigi}{\sigma_i}
\nuc{\sigk}{\sigma_k}
\nuc{\siga}{\sigma_a}
\nuc{\sigb}{\sigma_b}
\nuc{\Sw}{S_w}
\nuc{\Swf}{S_{w,f}}
\nuc{\Swb}{S_{w,b}}

\nuc{\Tb}{\bar T}
\nuc{\tn}{t_n}
\nuc{\tr}{t_r}
\nuc{\ti}{t_i}
\nuc{\ticl}{t_{i,l}}
\nuc{\trcm}{t_{r,m}}
\nuc{\twnm}{t_1^{n_m}}
\nuc{\twnl}{t_1^{n_l}}
\nuc{\twclnl}{t_{1,l}^{n_l}}
\nuc{\trmw}{t_{r-1}}
\nuc{\trpw}{t_{r+1}}
\nuc{\thehw}{\hat\theta_1}
\nuc{\theho}{\hat\theta_0}
\nuc{\thek}{\theta_k}
\nuc{\thej}{\theta_j}
\nuc{\thea}{\theta_a}
\nuc{\theb}{\theta_b}
\nuc{\theab}{\theta_{ab}}
\nuc{\thekj}{\theta_{kj}}
\nuc{\dtri}{\tr - \ti}
\nuc{\dtrcmicl}{\trcm - \ticl}
\nuc{\dTtr}{T - \tr}
\nuc{\dTti}{T - \ti}
\nuc{\dTticl}{T - \ticl}
\nuc{\dtr}{\trpw-\tr}
\nuc{\dtrcm}{t_{r+1,m}-\trcm}
\nuc{\tauwnicj}{\tau_1^{\nicj}}
\nuc{\tauicj}{\tau_{i,j}}
\nuc{\tauicjcl}{\tau_{i,j,l}}

\nuc{\ui}{u_i}
\nuc{\uk}{u_k}
\nuc{\uj}{u_j}
\nuc{\uij}{u_{ij}}
\nuc{\ukj}{u_{kj}}
\nuc{\uab}{u_{ab}}
\nuc{\ubi}{\bar u_i}
\nuc{\ubj}{\bar u_j}
\nuc{\Ui}{U_i}
\nuc{\Uj}{U_j}
\nuc{\Uij}{U_{ij}}
\nuc{\Ubi}{\bar U_i}
\nuc{\Ubj}{\bar U_j}

\nuc{\vi}{v_i}
\nuc{\vj}{v_j}
\nuc{\vij}{v_{ij}}
\nuc{\vbi}{\bar v_i}
\nuc{\vbj}{\bar v_j}
\nuc{\Vi}{V_i}
\nuc{\Vj}{V_j}
\nuc{\Vij}{V_{ij}}
\nuc{\Vbi}{\bar V_i}
\nuc{\Vbj}{\bar V_j}

\nuc{\wt}{w_t}
\nuc{\wtmw}{w_{t-1}}
\nuc{\wtpw}{w_{t+1}}
\nuc{\wicj}{w_{i,j}}
\nuc{\wicjclcm}{w_{i,j,l,m}}
\nuc{\wbt}{w_{b,t}}
\nuc{\wbtpw}{w_{b,t+1}}
\nuc{\wb}{w_b}

\nuc{\Xhat}{\hat X}

\nuc{\xt}{x_t}
\nuc{\xtpw}{x_{t+1}}
\nuc{\xtmw}{x_{t-1}}
\nuc{\xo}{x_0}
\nuc{\xw}{x_1}
\nuc{\xto}{x_2}
\nuc{\xT}{x_T}

\nuc{\xTt}{x^T_t}
\nuc{\xtt}{x^t_t}
\nuc{\xtmwt}{x^{t-1}_t}

\nuc{\xtil}{\tilde{x}}
\nuc{\xtilt}{\tilde{x}_t}
\nuc{\xtilu}{\tilde{x}_u}
\nuc{\xtiltgs}{\tilde{x}_{t|s}}
\nuc{\xtilugs}{\tilde{x}_{u|s}}
\nuc{\xtiltgt}{\tilde{x}_{t|t}}
\nuc{\xtiltgT}{\tilde{x}_{t|T}}
\nuc{\xtilugT}{\tilde{x}_{u|T}}
\nuc{\xtiltgtmw}{\tilde{x}_{t|t-1}}
\nuc{\xhat}{\hat{x}}
\nuc{\xhatt}{\hat{x}_t}
\nuc{\xhattgs}{\hat{x}_{t|s}}
\nuc{\xhattgtmw}{\hat{x}_{t|t-1}}
\nuc{\xhattt}{\hat{x}_t^t}
\nuc{\xhattgt}{\hat{x}_{t|t}}
\nuc{\xhattgT}{\hat{x}_{t|T}}
\nuc{\xhattmwgtmw}{\xhat_{t-1|t-1}}

\nuc{\xht}{\xh_t}
\nuc{\xh}{\hat{x}}
\nuc{\Xh}{\hat{X}}

\nuc{\yt}{y_t}
\nuc{\yr}{y_r}
\nuc{\ywn}{y_1^n}
\nuc{\ywT}{y_1^T}
\nuc{\yws}{y_1^s}
\nuc{\Yp}{Y_p}
\nuc{\Yf}{Y_f}
\nuc{\ytpw}{y_{t+1}}
\nuc{\ytmw}{y_{t-1}}
\nuc{\zt}{z_t}
\nuc{\ztpw}{z_{t+1}}
\nuc{\zit}{\zeta_t}
\nuc{\zitp}{\zeta_{t+1}}
\nuc{\clrb}[1]{{\color{blue}{#1}}}
\nuc{\clrr}[1]{{\color{red}{#1}}}

\nuc{\wpw}{\mbox{w.p.1.}}

\dmo{\T}{T}
\nuc{\bP}{\mathbb{P}}
\nuc{\cF}{\mathcal{F}}
\nuc{\cHtm}{\cH_{t-}}
\nuc{\cHsm}{\cH_{s-}}
\nuc{\cHom}{\cH_{0-}}
\renuc{\cI}{\mathbf{1}}

\nuc{\intitm}{\int_{-\infty}^{t-}}

\nuc{\umck}{unit-mass causal kernel}

\nuc{\limdo}{\lim_{\delta\to0}}
\nuc{\limdoo}{\lim_{\delta\to0}\wo\delta}

\renuc{\thehw}{\hat\theta_T^{(1)}}
\nuc{\wnt}{\mathcal{I}_{[0,\infty)}(t)}

\nuc{\WkTo}{W_{k,T}^0}
\nuc{\Sigthek}{\Sig_{\thek}}
\nuc{\Sigalpk}{\Sig_{\alpk}}
\nuc{\Sigck}{\Sig_{\ck}}

\nuc{\chij}{\chi_j}
\nuc{\phib}{\bar\phi}
\nuc{\phibo}{\bar\phi^{(0)}}
\nuc{\Mt}{M_t}
\nuc{\lamtilh}{\tilde\lam^{(h)}}
\nuc{\lamtilw}{\tilde\lam^{(1)}}
\nuc{\lamtilo}{\tilde\lam^{(0)}}
\nuc{\cw}{c^{(1)}}
\nuc{\alpw}{\alpha^{(1)}}
\nuc{\co}{c^{(0)}}
\nuc{\ch}{c^{(h)}}
\nuc{\chw}{\hat c_T^{(1)}}
\nuc{\csw}{c_*^{(1)}}

\nuc{\chijph}{\chi_{j'}^{(h)}}
\nuc{\chitilh}{\tilde\chi^{(h)}}
\nuc{\chih}{\hat\chi}
\nuc{\chitiljh}{\tilde\chi_j^{(h)}}
\nuc{\chitiljph}{\tilde\chi_{j'}^{(h)}}
\nuc{\chitilw}{\tilde\chi^{(1)}}
\nuc{\chitilwT}{\tilde\chi^{(1)\top}}
\nuc{\chitiljw}{\tilde\chi_j^{(1)}}
\nuc{\chitilo}{\tilde\chi^{(0)}}
\nuc{\chitiljo}{\tilde\chi_j^{(0)}}

\nuc{\alphw}{\hat\alpha_T^{(1)}}
\nuc{\alpsw}{\alpha_*^{(1)}}

\nuc{\alphwT}{\hat\alpha_T^{(1)\top}}

\nuc{\qjh}{q_j^{(h)}}
\nuc{\qjph}{q_{j'}^{(h)}}
\nuc{\qjw}{q_j^{(1)}}
\nuc{\qjo}{q_j^{(0)}}
\nuc{\qw}{q^{(1)}}
\nuc{\Ntilu}{\tilde N_u}
\nuc{\Ntilt}{\tilde N_t}
\nuc{\theh}{\theta^{(h)}}
\nuc{\thew}{\theta^{(1)}}
\nuc{\thewT}{\theta^{(1)\top}}

\nuc{\Thew}{\Theta^{(1)}}

\nuc{\NtilT}{\tilde N_T}
\nuc{\RTh}{R_T^{(h)}}
\nuc{\RTo}{R_T^{(0)}}
\nuc{\RTw}{R_T^{(1)}}

\nuc{\chihTh}{\hat\chi_T^{(h)}}
\nuc{\chihThT}{\hat\chi_T^{(h)\top}}
\nuc{\chihTw}{\hat\chi_T^{(1)}}
\nuc{\chihTwT}{\hat\chi_T^{(1)\top}}

\nuc{\STh}{S_T^{(h)}}
\nuc{\STw}{S_T^{(1)}}
\nuc{\STo}{S_T^{(0)}}

\nuc{\Nh}{\hat N_T}
\nuc{\GTh}{G_T^{(h)}}
\nuc{\GTo}{G_T^{(0)}}
\nuc{\GTw}{G_T^{(1)}}

\nuc{\gTw}{g_T^{(1)}}

\nuc{\sTh}{s_T^{(h)}}
\nuc{\sTw}{s_T^{(1)}}
\nuc{\sTwT}{s_T^{(1)\top}}
\nuc{\sTo}{s_T^{(0)}}

\nuc{\LamhT}{\hat\Lam_T}

\nuc{\Lamh}{\hat\Lambda}
\nuc{\lamh}{\lam^{(h)}}
\nuc{\chijh}{\chi_j^{(h)}}
\nuc{\chio}{\chi^{(0)}}
\nuc{\RTho}{R_T^{(h,0)}}
\nuc{\alpo}{\alpha^{(0)}}
\renuc{\Rsh}{R_*^{(h)}}
\nuc{\Rsw}{R_*^{(1)}}
\nuc{\Rso}{R_*^{(0)}}
\nuc{\Rsho}{R_*^{(h,0)}}
\nuc{\Rswo}{R_*^{(1,0)}}
\nuc{\muh}{\mu^{(h)}}
\nuc{\muhT}{\mu^{(h)\top}}
\nuc{\muo}{\mu^{(0)}}
\nuc{\muoT}{\mu^{(0)\top}}
\nuc{\muw}{\mu^{(1)}}
\nuc{\muwT}{\mu^{(1)\top}}
\nuc{\qh}{q^{(h)}}
\nuc{\qbh}{\bar q^{(h)}}
\nuc{\qbo}{\bar q^{(0)}}
\nuc{\qo}{q^{(0)}}
\nuc{\ssh}{s_*^{(h)}}
\nuc{\Nv}{N_v}

\nuc{\lamo}{\lam^{(0)}}
\nuc{\lamw}{\lam^{(1)}}
\nuc{\alpjo}{\alpha_j^{(0)}}
\nuc{\alpjw}{\alpha_j^{(1)}}
\nuc{\betw}{\beta^{(1)}}

\nuc{\Lto}{L_2[0,\infty)}
\nuc{\gw}{g^{(1)}}
\nuc{\gbw}{\bar g^{(1)}}

\nuc{\chihT}{\hat\chi_T}
\nuc{\alphT}{\hat\alpha^{(h)\top}}
\nuc{\alphTw}{\hat\alpha_T^{(1)}}
\nuc{\chTw}{\hat c_T^{(1)}}

\nuc{\wnPw}{\bm{1}_{P^{(1)}}}
\nuc{\wnPh}{\bm{1}_{P^{(h)}}}
\nuc{\alpsh}{\alpha_*^{(h)}}
\nuc{\ssw}{s_*^{(1)}}
\nuc{\VhTh}{\hat V_T^{(h)}}
\nuc{\VhTw}{\hat V_T^{(1)}}
\nuc{\RTwo}{R_T^{(1,0)}}

\nuc{\Sigh}{\Sigma^{(h)}}
\nuc{\Sigw}{\Sigma^{(1)}}
\nuc{\Sigo}{\Sigma^{(0)}}

\nuc{\phio}{\phi^{(0)}}
\nuc{\phiw}{\phi^{(1)}}
\nuc{\Dphi}{\Delta\phi}
\nuc{\RThh}{R_T^{(h,h)}}
\nuc{\chihp}{\chi^{(h')}}
\nuc{\chitilhp}{\tilde\chi^{(h')}}
\nuc{\phisw}{\phi^{(1)}_*}
\nuc{\Gamsw}{\Gam_*^{(1)}}
\nuc{\Gamsh}{\Gam_*^{(h)}}
\nuc{\Gamso}{\Gam_*^{(0)}}
\nuc{\Fphih}{F_\phi^{(h)}}
\nuc{\Fphiw}{F_\phi^{(1)}}
\nuc{\Fphio}{F_\phi^{(0)}}
\nuc{\Sphih}{S_\phi^{(h)}}
\nuc{\Sphiw}{S_\phi^{(1)}}
\nuc{\Sphio}{S_\phi^{(0)}}
\nuc{\phih}{\phi^{(h)}}
\nuc{\sqintT}{\wo{\sqrt{T}}\intoT}
\nuc{\intT}{\wo{T}\intoT}
\nuc{\intoTs}{\int_{(0,T]^2}}
\nuc{\phibw}{\bar\phi^{(1)}}
\nuc{\phibws}{\bar\phi^{(1)}_*}

\dmo{\WT}{W_{j}}
\nuc{\WsT}{W^*_T}
\nuc{\intoum}{\int_0^{u_-}}
\nuc{\qfpf}{q\star f\cdot \Dphi\star f}
\nuc{\qfpfu}{q\star f(u)\Dphi\star f(u)}

\nuc{\intTi}{\int_T^\infty}
\nuc{\intui}{\int_u^\infty}
\nuc{\intotT}{\int_0^{\tau T}}
\nuc{\intoTmu}{\int_0^{T-u}}

\nuc{\Sef}{S_{ef}}

\renuc{\lamw}{\lam_1}
\renuc{\alpw}{\alpha_1}
\nuc{\chiw}{\chi_1}
\nuc{\Lw}{L^1[0,\infty)}
\renuc{\Lt}{L^2[0,\infty)}
\nuc{\Lwp}{L^1_+[0,\infty)}
\nuc{\Lwn}[1]{\Ver{#1}_{L^1}}
\nuc{\Lin}[1]{\Ver{#1}_{L^\infty}}
\nuc{\Lpn}[1]{\Ver{#1}_{L^p}}
\nuc{\Ltn}[1]{\Ver{#1}_{L^2}}
\renuc{\Li}{L^\infty[0,\infty)}
\nuc{\lamtil}{\tilde\lam}
\nuc{\chitil}{\tilde\chi}
\renuc{\phio}{\phi_0}
\renuc{\phiw}{\phi_1}
\renuc{\lamo}{\lam_0}
\renuc{\lamtilw}{\tilde\lam_1}

\nuc{\intotm}{\int_0^{t-}}
\nuc{\intium}{\int_{-\infty}^{u-}}
\nuc{\intiu}{\int_{-\infty}^{u}}
\nuc{\theT}{\theta^\top}
\nuc{\chihTT}{\hat\chi_T^\top}
\renuc{\chihT}{\hat\chi_{_T}}
\renuc{\chihTT}{\hat\chi_T^\top}
\renuc{\chihTw}{\hat\chi_{_T}}
\renuc{\chihTwT}{\hat\chi_T^\top}
\nuc{\GT}{G_T}
\nuc{\gT}{g_{_T}}
\nuc{\RT}{R_T}
\nuc{\sT}{s_{_T}}
\renuc{\ch}{\hat c}
\renuc{\theh}{\hat\theta}
\renuc{\chio}{\chi_0}
\renuc{\alpo}{\alp_0}
\nuc{\bw}{\mathbbm{1}}
\nuc{\Rs}{R_*}
\nuc{\muT}{\mu^\top}
\nuc{\qb}{\bar q}
\renuc{\phibo}{\bar\phi_0}
\nuc{\alps}{\alpha_*}
\nuc{\cs}{c_*}
\renuc{\co}{c_0}

\nuc{\Creg}{C_{reg}}
\nuc{\kap}{\psi}
\nuc{\kapc}{\check\psi}
\nuc{\intiom}{\int_{-\infty}^{0-}}

\nuc{\snwi}{\sum_{n=1}^\infty}

\nuc{\chijp}{\chi_{j'}}
\nuc{\chitilj}{\tilde\chi_j}
\nuc{\intoTm}{\int_0^{T-}}

\nuc{\VhT}{\hat V_T}
\nuc{\thes}{\theta_*}
\renuc{\chitilo}{\tilde\chi_0}
\nuc{\Dchio}{\Delta\chi_0}
\nuc{\kapb}{\bar\kap}

\nuc{\BaT}{B^\alp_T}
\nuc{\BcT}{B^c_T}
\nuc{\ot}{^{\otimes 2}}

\nuc{\Sigoth}{\Sig_0^\theta}
\nuc{\Sigoalp}{\Sig_0^\alpha}
\nuc{\Sigcrb}{\Sig_{CRB}}
\nuc{\sigoc}{\sig_0^c}
\renuc{\Sigo}{\Sig_0}
\nuc{\Gs}{G_*}

\nuc{\xitil}{\tilde\xi}
\nuc{\Dchi}{\Delta\chi}
\nuc{\Dchij}{\Delta\chi_j}

\nuc{\Gams}{\Gam_*}
\nuc{\pt}{pseudo-true}
\nuc{\Mr}{Martingale representation}
\nuc{\mr}{martingale representation}
\nuc{\phis}{\phi_*}
\nuc{\phij}{\phi_j}

\dmo{\Sr}{S}
\nuc{\Sqj}{S^q_j}
\nuc{\Sq}{S^q}
\renuc{\Sphio}{S^\phi_0}
\nuc{\Sphis}{S^\phi_*}
\nuc{\Skap}{S^\kap}
\nuc{\sqT}{\sqrt{T}}
\nuc{\phibs}{\bar\phi_*}

\nuc{\DW}{\Delta W_T}
\nuc{\DWj}{\Delta W_{j,T}}
\nuc{\BDGi}{Burkholder-Davis-Gundy inequality}

\nuc{\Vs}{V^{T,u}_s}
\nuc{\VTu}{V^{T,u}}
\nuc{\intosum}{\int_0^{su-}}
\nuc{\intosu}{\int_0^{su}}

\nuc{\Wtil}{\tilde W}
\renuc{\ha}{h^\alp}
\nuc{\Dphio}{\Delta\chi_0}

\renuc{\muh}{\mu_h}
\nuc{\muha}{\mu_{h}^{\alpha}}
\nuc{\xitilh}{\tilde\xi_h}
\renuc{\chitilh}{\chitil_h}

\nuc{\Sigsth}{\Sigma_*^\theta}
\nuc{\Sigsalp}{\Sigma_*^\alp}
\nuc{\sigsc}{\sig_*^c}
\nuc{\Sigs}{\Sig_*}

\nuc{\chihh}{\chi_h}
\nuc{\chis}{\chi_*}
\nuc{\chitilha}{\tilde\chi_h^\alp}

\nuc{\MtilxihT}{\tilde M^\xi_{h,T}}
\nuc{\MtilxiT}{\tilde M^\xi_{T}}

\nuc{\dome}{\delta_\omega}
\nuc{\dt}{\delta_t}
\nuc{\Dphib}{\overline{\Dphi}}

\renuc{\pro}[1]{{\it Proof. }#1\hfill$\square$}

\renewcommand{\theequation}{\arabic{section}.\arabic{equation}}

\newtheorem{remark}{Remark}

\begin{document}

\begin{frontmatter}

\title{Hawkes Identification with a Prescribed Causal Basis: Closed-Form Estimators and  Asymptotics
\thanksref{footnoteinfo}} 

\author[UoM]{Xinhui Rong}\ead{xinhui.rong@unimelb.edu.au},    
\author[UoM]{Girish N. Nair}\ead{gnair@unimelb.edu.au}              
\thanks[footnoteinfo]{
Part of this manuscript has been presented at the 2026 American Control Conference. }

\address[UoM]{Department of Electrical and Electronic Engineering, 
the University of Melbourne, Australia}  

\begin{keyword}                           
System Identification, Hawkes Processes, Stochastic Systems, Asymptotic Analysis.
\end{keyword}                             

\begin{abstract}                          
Driven by the recent surge in neural-inspired modeling, 
point processes have gained significant traction in systems and control. 
While the Hawkes process is the standard model 
for characterizing random event sequences with memory, 
identifying its unknown kernels is often hindered by nonlinearity. 
Approaches using prescribed basis kernels have emerged 
to enable linear parameterization, 
yet they typically rely on iterative likelihood methods 
and lack rigorous analysis under model misspecification. 
This paper justifies a closed-form Least Squares 
identification framework for Hawkes processes with prescribed kernels. 
We guarantee estimator existence 
via the almost-sure positive definiteness of the empirical Gram matrix 
and prove convergence to the true parameters under correct specification, 
or to well-defined pseudo-true parameters under misspecification. 
Furthermore, we derive explicit Central Limit Theorems for both regimes, 
providing a complete and interpretable asymptotic theory. 
We demonstrate these theoretical findings through comparative numerical simulations.
\end{abstract}

\end{frontmatter}

\mysec{Introduction}\label{sec:intro}
The past few decades have witnessed a paradigm shift in the monitoring and understanding of event-based processes, driven by a surge in neural-inspired technologies and the unprecedented availability of event data. This shift spans diverse disciplines: from computational neuroscience \cite{Bial97}, where information is encoded in the precise timing of neural spikes rather than voltage magnitude; to genomics \cite{Cars10}, where transcription events occur at specific genomic positions; to high-frequency finance \cite{Bacr15,Embr11}, where trade timing is critical for inferring market structure; and social media analysis \cite{Zhou13}, where the causality of tweet and retweet events reveals social network topology. Consequently, these random-event frameworks have recently found key applications in control and signal processing, including event-triggered state estimation \cite{Shi15}, event-based vision \cite{Gall22}, and multitarget radar tracking \cite{G-Fern23}.

Point processes \cite{Dale03} provide a precise mathematical framework for modelling these event-based phenomena, drawing on over sixty years of theoretical development across statistics, biometrics, econometrics, etc. Within this framework, the Hawkes process \cite{Hawk71} stands out as a causal, self-exciting model, serving as the point-process counterpart to autoregressive models. 

Traditional maximum likelihood estimators (MLEs), 
implemented via direct numerical optimization \cite{Ozak79} or the expectation-maximization (EM) algorithm \cite{Veen08,Lewi11,Godo20} are available for the Hawkes process. 
However, the identification of Hawkes processes remains an active research area, even for the standard linear form. This challenge arises because, although the intensity is linear with respect to the memory regressor, the regressor itself is defined by a convolution of past events with a deterministic Hawkes Impulse Response (HIR) that typically depends nonlinearly on the parameters, e.g., 
the exponents. 
The number of nonlinear parameters grows quadratically as the network dimension, 
rendering the iterative likelihood methods intractable \cite{DFon14}. 

Non-parametric Hawkes modelling \cite{Kirc17,Bacr12,Acha17} circumvents the challenges of nonlinear parameter estimation but typically relies on binning the process in either the time or frequency domain, which introduces tuning sensitivity and inevitable information loss. 
An alternative approach avoids binning by modelling the HIR as a linear combination of prescribed, causal basis functions. In this framework, events are filtered rather than bin-counted, preserving the continuous-time (CT) nature of the data without information loss while rendering the model linear in all parameters.

Using a prescribed basis (e.g., Hawkes-Laguerre \cite{Ogat82,Godo20}) yields significant computational advantages. The EM updates become closed-form \cite{Godo20,Lewi11}. More importantly, the linearity enables a CT Least-Squares (LS) formulation \cite{Rong23,Bacr20,Hans15,Reyn10,Meno18}. This admits potential closed-form estimates, provided the empirical Gram matrix is positive definite, and offers a much more straightforward framework for incorporating sparsity or other parameter constraints. 

The LS approach naturally raises three fundamental questions: 
(1) Under what conditions is the empirical Gram matrix positive definite with probability one (w.p.1), thereby guaranteeing the existence and uniqueness of the closed-form LS estimate? (2) If the true HIR lies within the span of the prescribed basis (correct specification), is the LS estimator consistent, and does it exhibit asymptotic normality? (3) In the inevitable scenario where the true HIR is misspecified by the prescribed basis, does the estimator converge to a well-defined pseudo-true parameter, and does a Central Limit Theorem (CLT) still hold?

In the conference version \cite{Rong26} of this paper, 
we partly answered the above questions, complementing the existing results. 
Unlike prior studies that either assume positive definiteness or guarantee it only with high probability \cite{Hans15}, in \cite{Rong26}, we established that the empirical Gram matrix is positive definite w.p.1 under minimal conditions. 
The second and third questions present significant challenges in Hawkes process identification, even within the well-studied MLE framework. Under correct model specification, asymptotic consistency and CLTs have been established for Hawkes MLEs \cite{Ogat78,Clin17,Kwan23}, and consistency has been shown for binned processes \cite{Kirc17}. In the context of CT LS, while finite-sample risk bounds exist for penalized estimators \cite{Bacr20,Reyn10}, we recently derived the first consistency result for the unpenalized estimator in \cite{Rong26}, relying on an ergodicity assumption. Regarding model misspecification, although general quasi-likelihood theory is well-established \cite{Whit82}, 
the asymptotic convergence of the CT LS estimators remained an open problem until \cite{Rong26}, where we proved convergence in probability to pseudo-true parameters under the ergodicity assumption. However, to date, no CLTs have been established for CT LS estimators under either correct or misspecified conditions.

In this paper, we substantially strengthen the results of \cite{Rong26} in two key aspects. First, we replace the general ergodicity assumption with explicit causal kernel moment conditions, underpinning the asymptotic analysis and thereby upgrading the convergence from `in probability' to `w.p.1'. Second, we establish CLTs for CT LS estimators under both correct model specification and misspecification. For the sake of completeness, we also recapitulate the preliminary results from \cite{Rong26}, including the existence conditions of the LS estimators. 
We focus on one-dimensional (not spatial), univariate (not network) linear Hawkes processes.

These theoretical findings substantiate the use of LS Hawkes identification with a prescribed basis, challenging the prevailing view that LS is merely a secondary alternative to MLE. Beyond these immediate contributions, our results open the door to several significant applications. The explicit nature of the LS estimator promises to enable real-time implementation with substantial computational advantages. Furthermore, the derivation of explicit pseudo-true parameters and asymptotic covariance paves the way for rigorous robustness analysis in future studies. Significantly, the established CLTs provide the essential theoretical foundation for developing Generalized Method of Moments (GMM) tests \cite{Hall05}, particularly for non-nested model comparisons \cite{Vuon89,Rive02} under misspecification.

The remainder of this paper is organized as follows. 
Section \ref{sec:pre} and Section \ref{sec:mod} review the necessary preliminaries on the general point processes and key lemmas for Hawkes processes. 
Section \ref{sec:id} formulates the CT LS problem, 
derives the closed-form estimators, and establishes their finite-time existence. 
Section \ref{sec:conv} analyzes the asymptotic behavior of the estimators, 
proving convergence to the true parameters under correct specification 
and to well-defined pseudo-true parameters under misspecification. 
Section \ref{sec:clt} derives the CLTs for both specification regimes, providing explicit and interpretable expressions for the asymptotic covariances.
Section \ref{sec:sim} presents a numerical study that computes \pt\ values, conducts an asymptotic robustness analysis of the \HL\ model, and validates the asymptotic properties of the LS estimator.
Section VIII contains conclusions and future work.

\mysec{Preliminaries}\label{sec:pre}
In this section, we introduce the mathematical characterization of 
point processes and review the fundamental notations required 
for our analysis. 

A point process is a process of random event times. 
Denote the strictly increasing sequence of event times 
(e.g., the timings of neural spikes) 
by $\{t_r\}_{r\in\bZ}$. 
The {\it counting process} $N_t$ represents a point process 
and is defined by 
\eq{
N_t\trieq N((-\infty,t])=\ssum{r\in\bZ}\cI_{t\geq t_r},
}
where $N(A)$ measures the number of events in a Borel set $A\subset\bR$ 
and $\cI_{t \in A}=\{\smat{1,&t\in A\\0, &t\notin A}$ is the indicator function. 
By definition, $N_t$ is a right-continuous non-decreasing step function 
that jumps by one at each event time.

To characterize the dynamics, 
it is convenient to work with the differential increment of $N_t$. 
Heuristically, the counting increment $dN_t = \sum_{t_r\leq t}\delta(t-t_r)dt$ 
can be interpreted as a train of Dirac impulses $\delta$ localized at the event times. 
This interpretation allows us to define filtering operations 
via \RSc. For a causal kernel function 
$g: [0,\infty)\mapsto\bR$ and an observation window $A\subset\bR$, 
the \RSc\ of the kernel with the counting process is defined as 
$\int_A g(t-u)dN_u = \sum_{t_r\in A} g(t-t_r)$. 

Throughout, we adopt the standard {\it orderliness} assumption:  
$\lim_{\tau\to0}\wo\tau\Pr[N([t,t+\tau))>1]=0, \forall t\in\bR$, 
which suggests that simultaneous events are not allowed in an infinitesimal interval. 
Let $\cHtm = \sigma\{N_s, s<t\}$ be the history/filtration of the counting process. 
The {\it (conditional) intensity function} $\lam$ is defined as 
\eqn{
\lam(t) 
	&= \lim_{\tau\to0}\wo\tau\E[N([t,t+\tau))|\cHtm],
}
the instantaneous probability rate of a future event, 
conditioned strictly on the past history. 
The intensity function $\lam$ uniquely characterizes a point process.

We need to introduce some basic notations. 
Let $g,f: \bR\to\bR$ be deterministic functions. 
We denote $\bar g(s)=\intii e^{-st}g(t)dt$ the Laplace transform (LT) of $g$ 
and $\bar g(\jw)$ is, therefore, the Fourier transform (FT) of $g$. 
We denote $g\star f(t) = \intii g(t-u)f(u)du$ the convolution of $f$ and $g$. 
We denote $g\star dN_t = \intitm g(t-u)dN_u$ the \RSc. 
We will also write $|g|\star dN_t = \intitm |g(t-u)|dN_u$. 
We denote $f^{\star n}(t)=f\star f\star\dotsm\star f(t)$ the $n$-fold convolution of $f$.  
These definitions also extend when $g: \bR\mapsto\bR^k$ is a vector of functions. 
When we put the absolute sign on a vector $g(t)=[g_1(t),\dotsm,g_P(t)]^\top$, 
we mean $|g(t)|=[|g_1(t)|,\dotsm,|g_P(t)|]^\top$. 
The $L^p$ norm is defined as $\Lpn{f}=\bra{\int_\bR |f(t)|^pdt}^{1/p},1\leq p<\infty$ 
and $\Lin{f}=\sup_{t\in\bR} |f(t)|$. 
We denote $L^p[0,\infty),1\leq p\leq\infty$ the $L^p$ space 
of functions $f: \bR\mapsto \bR$ such that $f(t)=0,t<0$ and 
their $L^p$ norm $\Lpn{f}$ exists. 
Let $\bbw P$ denote a $P$-dimensional vector of $1$'s. 
The notation $o_p(\bbw P)$ represents a sequence of $P$-dimensional random vectors that converges to zero in probability. 
The notation a.e. stands for almost everywhere, 
the notation $\convp$ stands for convergence in probability, 
and $X_T\RA\cN(0,\Sigma)$ means that $X_T$ converges in distribution to a zero-mean Gaussian random variable with covariance matrix $\Sigma$.

\mysec{Hawkes Modelling}\label{sec:mod}
The Hawkes process is the most widely used model 
for history-dependent event sequences, 
serving as the point-process counterpart to the autoregressive model in time series. 
This section reviews the Hawkes intensity and the fundamental properties 
that establish the basis for our analysis and 
motivate the assumptions adopted in this work.

\subsection{The Hawkes Intensity}
A Hawkes process is a self-exciting point process 
uniquely characterized by the intensity function
\begin{subequations}\label{eq:lam}%
\begin{align}
	&\lamo(t) &=& c_0 + \chio(t)\\%
	&\chio(t) &=& \phio\star dN_t=\intitm\phio(t-u)dN_u\\
\mbox{with } &\phio(t)&\geq&0, \quad\phio\in\Lw\cap\Li, \\
\mbox{and } &\Gam &=& \Lwn{\phio}<1, c_0>0.
\end{align}
\end{subequations}
$c_0$ is the deterministic {\it background rate}, 
$\chio$ is a stochastic {\it memory process}, 
which is the self-exciting component of the intensity, 
$\phio$ is the deterministic {\it Hawkes impulse response} (HIR), 
characterizing the additive increase in the intensity 
induced by each past event, 
and $\Gam$ is the {\it branching ratio}. 

The conditions on the HIR $\phio$ are to ensure 
the almost-sure positivity and boundedness of the intensity $\lamo$ 
as well as the stationarity, which underpins the key properties, 
to be introduced next. 
The interpolation inequality \cite{Foll99} ensures $\Lpn{\phio}\leq\Lwn{\phio}^{1/p}\Lin{\phio}^{1-1/p}<\infty$, for any $1< p<\infty$. 
By definition, the Hawkes intensity $\lamo$ 
is left-continuous with right limits, 
so that it is an $\cHtm$-predictable stochastic process \cite{Brem81}.

Under a Hawkes intensity \eqref{eq:lam}, 
the deterministic {\it Hawkes resolvent} is defined as
\eqn{\label{eq:kap}
\kap(t) = \sum_{n=1}^\infty \phio^{\star n}(t), \quad t\geq0.
}
The existence of the Hawkes resolvent $\kap$ is guaranteed by the condition $\Gam\in(0,1)$ and it is straightforward to show 
that $\Lpn{\kap}<\infty$ for all $1\leq p\leq\infty$ recursively using the Young's convolutional inequality 
and that $\Lwn{\kap}=\snwi\Lwn{\phio}=\frac{\Gam}{1-\Gam}$ 
by recursively using the identity $\Lwn{\phio^{\star2}}=\Lwn{\phio}^2$. The Hawkes resolvent has LT
\eqn{\label{eq:kapb}
\kapb(s) = \frac{\phibo(s)}{1-\phibo(s)},
}
and is closely related to the higher-order statistics 
of the counting process and our asymptotic analysis.

\subsection{Key Lemmas}
Here, we summarize key properties of Hawkes processes,  
including stationarity and ergodicity of the counting process, 
moments and cumulants, and martingale dynamics. 
These results are either standard or follow immediately 
from known properties, with short derivations included for completeness.

\Blem{lem:sta} {\bf Stationarity.}\cite{Brem96} 
There exists a unique strictly stationary distribution 
for the counting process $N_t$, characterized by (\ref{eq:lam}).   
\Elem

\Blem{lem:erg}{\bf Ergodicity. }
A stationary Hawkes counting process with (\ref{eq:lam}) is ergodic\footnote{An ergodic counting process $N$ has a trivial invariant $\sigma$-algebra under the measure-preserving shift operations. See \cite[Chapter 12]{Dale08} for full details.}.
\Elem
\pro{
A stationary cluster process is ergodic if its cluster center is \cite[Proposition 12.3.IX]{Dale08}. A Hawkes process with (\ref{eq:lam}) is  
a Poisson cluster process \cite{Hawk74} 
whose cluster center follows a time-invariant Poisson process.  
Since the time-invariant Poisson process is ergodic \cite[Exercise 12.3.1]{Dale08}, 
the resulting counting process is ergodic. 
}

The following Lemmas \ref{lem:1mom}-\ref{lem:hmom} summarize 
properties of the cumulants of the counting increments. 
We require the explicit formulae for the first and second cumulants 
and the existence of all higher-order integrated moments. 

\Blem{lem:1mom}{\bf First order statistics. }\cite{Hawk71}
For a stationary Hawkes process with (\ref{eq:lam}), 
the expected counting increment is given by\footnote{Throughout the paper, the expectation $\E$, the probability $\Pr$, the variance $\Var$, and the covariance $\Cov$ are taken with respect to the unique stationary probability measure of the counting process $N$ defined by the intensity \eqref{eq:lam}.}
\eq{
\E[dN_t]/dt =\E[\lamo(t)]=\E[\lamo(0)]= \Lam ,
}
where $\Lam = \frac{c_0}{1-\Gam}$ is the {\it expected rate}. 
\Elem

\Blem{lem:2mom}{\bf Second order statistics. }
For a stationary Hawkes process with (\ref{eq:lam}), \\
(a) the covariance density of $N$ is given by
\eq{
C(u-v) = \frac{\E[d\Nu dN_{v}] - \E[d\Nu]\E[dN_{v}]}{dudv} = C(v-u). 
}
(b) Further, the Bartlett's spectrum, which is the FT of $C$ and given by\footnote{We write $\bar{C}(\omega)$ instead of $\bar{C}(\jmath\omega)$ to denote that the spectrum is real.}
\eqn{\label{eq:Cb}
\bar C(\omega) = \frac{\Lam}{|1-\phib(\jmath\omega)|^2},
}
exists, is real, and is strictly positive at any $\omega\in\bR$.\\
(c) The covariance density $C(\tau)$ contains a singular spike at the origin 
and a bounded regular part as follows. 
\eqn{\label{eq:C2}
C(\tau) &= \Lam\delta(\tau) + \Creg(\tau)\\
\Creg(\tau) &= \Lam(\kap(\tau)+\kapc(\tau)+\kap\star\kapc(\tau)),
}
where $\delta(\tau)$ is the Dirac delta function centered at $0$, 
$\kapc(\tau)=\kap(-\tau)$, 
$\Creg$ is symmetric, non-negative, and bounded,  
and $\Lwn{C}=\frac{\Lam}{(1-\Gam)^2}<\infty$. 
\Elem

\pro{
Results (a,b) are in \cite{Hawk71} and \cite{Bacr15}. 
The existence and strict positivity of $\bar C(\omega)$ 
follows from $|\phib(\jw)|\leq\Gam<1$. 
Result (c) can be found in \cite{Bacr12} 
or be directly derived from the FT \eqref{eq:Cb}. 
The symmetry of $\Creg$ is obvious from its definition. 
The boundedness of $\Creg$ follows  
by noticing $\Lin{\kap}=\Lin{\kapc}$ and using Young's inequality to get $\Lin{\kap\star\kapc}\leq\Lin{\kap}\Lwn{\kap}<\infty$. 
$\Lwn{C}=\Lam(\Lwn{\delta}+2\Lwn{\kap}+\Lwn{\kap}^2)=\Lam(1+\frac{\Gam}{1-\Gam})^2=\frac{\Lam}{(1-\Gam)^2}$. 
}

\Blem{lem:hmom}{\bf Higher-order statistics. }\cite{Jova15} 
For a stationary Hawkes process with \eqref{eq:lam}, 
the $n$-th integrated moment 
is given by 
\eq{
\int_{[\tau_1,\dotsm,\tau_{n-1}]^\top\in\bR^{n-1}}\E[dN_tdN_{t+\tau_1}\dotsm dN_{t+\tau_{n-1}}]=K_ndt,
} 
where the constant $K_n\in(0,\infty)$ is bounded for all $n\in\bN$ and admits a recursive expression regarding $\Lam$ and $\Lwn{C}$ 
given in \cite[Section IV]{Jova15}. 
\Elem

The predictability of the Hawkes intensity $\lamo$ 
and the finite expected rate under $\Gam<1$ 
ensures the existence of the unique Doob-Meyer decomposition 
to construct a martingale. 
We define the {\it truncated counting process} 
\eq{
\Ntilt=N((0,t]),
} and $M_t = \Ntilt-\intot\lamo(u)du, dM_t = dN_t - \lamo(t)dt, t>0$.

\Blem{lem:mar}{\it Martingales dynamics. } \cite[Chapter II]{Brem81}
For a stationary Hawkes process with (\ref{eq:lam}), 
\nlist{
\ita $M_t$ is an $\cHtm$-martingale, i.e. for any $t\geq 0$, 
$\E[\Ntilt - \intot\lamo(u)du|\cHom]=0$.
\itb If $h(t)$ is an $\cHtm$-predictable process such that 
$\E[\intot |h(u)|\lamo(u)du]<\infty, t\geq0$, 
then $\intot h(u)dM_u$ is an $\cHtm$-martingale, 
i.e. for any $t\geq0$, $\E[\intot h(t)d\Ntilt - \intot h(t)\lamo(t)dt|\cH_{0-}]=0$.
}
\Elem

\mysec{Least-Squares Identification for Hawkes}\label{sec:id}
In this section, we impose the fundamental modelling assumptions, 
formulate the LS identification problem, 
and detail the closed-form estimators originally derived 
in our conference paper \cite{Rong26}. 
These results are restated here to ensure the exposition is self-contained 
and to establish the necessary framework for the asymptotic analysis 
in subsequent sections.

\subsection{Truncated Observation}
While the theoretical process extends to the infinite past, 
practical applications are limited to finite observation intervals. 
We therefore work with the truncated counting process $\Ntilt$, 
observed in a deterministic time period $(0,T]$. 
For the sake of simplicity, we assume that the counting process is stationary. 

\Bass{A1}{\bf Stationarity and Observation.}
The observed counting process $\Ntilt=N((0,t])$ is a truncation of 
the stationary counting process $N_t$ 
with the unknown true intensity \eqref{eq:lam}, 
over the time interval $t\in(0,T]$. 
\Eass

Under \ref{A1}, the full counting process $N_t$ is stationary and ergodic, 
satisfying all properties in 
Lemmas \ref{lem:sta}-\ref{lem:hmom}. 
The truncated counting process $\Ntilt$ satisfies Lemma \ref{lem:mar} 
and its increment $d\Ntilt$ preserves stationarity. 

\subsection{The Candidate Intensity and its Truncation}
The identification objective is to estimate both the background rate $c_0$ 
and the HIR $\phio$. 
However, the nonlinear parameterization of $\phio$ typically poses significant estimation challenges. 
To circumvent this, it is a common practice to approximate $\phio(t)=\sum_j\alpj q_j(t)$ 
using a linear combination of prescribed causal basis functions $q_j$, 
e.g. the Hawkes-Laguerre model \cite{Godo20}. 
This leads to the definition of the following linear candidate intensity
\begin{subequations}\label{eq:lam1}
\eqn{
\lamw(t;\theta) 
	&= c + \alp^\top\chi(t), \quad t\in\bR\\
\chi(t) &= q\star dN_t = \intitm q(t-u)dN_u\in\bR^P,\label{eq:lam13}
}
\end{subequations}
where $c$ is the candidate background rate, 
$\alp=[\alp_1,\alp_2,\dotsm,\alp_P]^\top$ is the weighting parameter 
with model order $P$, 
$\theta = [\alp^\top,c]^\top\in\bR^{P+1}$ is the parameter vector, 
and $\chi(t)=[\chi_1(t),\dotsm,\chi_P(t)]^\top$ is a vector of the candidate memory regressor, 
where the deterministic $q(t)=[q_1(t),\dotsm,q_P(t)]^\top$ 
is the vector of prescribed/user-defined {\it unit mass causal kernels} (UMCKs) 
with each UMCK $q_j\in\Lw\cap\Li$, and $q_j(t)\geq0$ and $\Lwn{q_j}=1$. 
We also define $\phiw(t) = \alp^\top q(t)$ as the candidate HIR. 

Since, again, the events before time $0$ are not observed, 
to identify the candidate intensity, we need to work with its truncated version
\begin{subequations}\label{eq:lam1til}
\eqn{
\lamtilw(t;\theta) &= c + \alp^\top\chitil(t),\quad t\in[0,T]\\
\chitil(t) &= q\star d\Ntilt\trieq \intotm q(t-u)dN_u\in\bR^P,\label{eq:lam1til2}
}
\end{subequations}
where $\chitil(t)\in\bR^P$ is the truncated memory regressor 
and we define the truncated \RSc\ as $g\star d\Ntilt = \intotm g(t-u)d\Ntilu$ 
of a causal function $g: [0,\infty)\mapsto\bR^k$ 
with the truncated counting process $\tilde N$ on $[0,t)$. 
To avoid confusion, we note that 
since $N$ and $\tilde N$ share the same increments on $(0,T]$, 
$g\star d\Ntilt=\intot g(t-u)dN_u = \intot g(t-u)d\Ntilu$. 
Subsequently, we will also write $|g|\star d\Ntilt=\intotm|g(t-u)|dN_u$. 
We note that the truncated processes lose stationarity.

Thanks to the prescribed UMCKs, given an observation $\tilde N$, 
the truncated memory regressor $\chitil$ is fully observed in $(0,T]$, 
leading to a linear parameterization of $\lamtilw$ in regard to the parameter $\theta$. For future use, we define 
\eqn{
\Dchi(t)&=\chi(t)-\chitil(t)\label{eq:Dchi}\\ 
\mbox{with } \Dchij(t) &= \chi_j(t)-\chitil_j(t),\quad j=1,\dotsm,P \nonumber\\
	&= \intiom q_j(t-u)dN_u>0.\label{eq:Dchij}
} 
We also define the truncated true (latent) memory process as 
$\chitilo(t) = \intotm\phio(t-u)dN_u$  
and $\Dchio(t) = \chio(t)-\chitilo(t) = \intiom\phio(t-u)dN_u>0$. 

The prescribed UMCKs need to satisfy the following identifiability condition 
which also guarantees a unique LS estimator to be developed next.

\Bass{A2}
{\bf Identifiable UMCKs. }
The candidate UMCKs satisfy the following conditions. 
\nlist{
\ita Density-like: $\qj\in\Lw\cap\Li$ with $\qj(t)\geq0$ and $\Lwn{q_j}=1$ for all $j$. 
\itb Finite-horizon affine independence: 
There exists a finite $t_0>0$, such that 
there is no pair of $x\neq0\in\bR^{P}$ and $d\in\bR$, 
such that $x^\top q(t)= d$,  a.e. on $t\in[0,t_0]$. 
\item[(*)] We define the {\it minimal affine-independence horizon} $T_0$ to be 
the infimum of such $t_0$ for which \ref{A2}(b) is satisfied.
}
\Eass

We note that \ref{A2}(a) is just a convenient rescaling of the HIR basis 
and that the interpolation inequality \cite{Foll99} ensures $\Lpn{q_j}<\infty$ for any $1\leq q\leq\infty$. 
\ref{A2}(b) strengthens ordinary linear independence in two respects:
(i) it enforces independence on the {\it finite} horizon $[0,t_0]$, 
and (ii) it requires affine independence.  
Equivalently, $\{1, q_1, \dotsm, q_P\}$ is linearly independent 
a.e. on $[0,t_0]$. 
These strengthenings ensure the existence and uniqueness of the LS estimator developed next. 
The assumption is mild and holds with $T_0=0$ for most prescribed UMCKs 
(e.g., exponential \cite{Hawk71} and Erlang bases \cite{Godo20}). 
If the candidate kernels are linearly independent but any one of them is constant on an initial interval, 
then $T_0$ equals the first time at which every kernel becomes time-varying. 
Since $\qj$'s are prescribed, 
$T_0$ is always known and can always be designed to be $0$.

\subsection{The Least-Squares Formulation}
Suppose we observe a truncated counting process with strictly increasing event times $t_1, t_2, \dots$ over a deterministic interval $(0,T]$. We define the truncated extended regressor as $\xitil(t) = [\chitil(t)^\top, 1]^\top \in \mathbb{R}^{P+1}$, where the appended constant accounts for the memoryless background rate. 
While the continuous-time least squares (LS) contrast, $J_T(\theta)$ given below, is well-established \cite{Hans15, Bacr20}, we decompose its key terms to yield a more interpretable LS estimator and streamline its asymptotic analysis. 
\eqn{\label{eq:ls}
J_T(\theta) &= \woT\intoT\lamtilw(t;\theta)^2dt - \frac2T\intoT\lamtilw(t;\theta)d\Nt\nonumber\\
	&= \theT \GT\theta - 2\theT\gT,
}
where  
$\GT= \woT\intoT\xitil(t)\xitil(t)^\top dt=\sqbra{\smat{\RT+\chihT\chihTT&\chihT\\\chihTT&1}}$ 
is the empirical Gram matrix 
and $\gT = [(\sT+\LamhT\chihT)^\top, \LamhT]^\top\in\bR^{P+1}$, 
where we define the scalar $\hat\Lam_T=\NtilT/T$ as the empirical rate, 
the vector $\chihT = \woT\intoT\chitil(t)dt\in\bR^P$ as the memory regressor mean, and
\eqn{
\RT &= \woT{\intoT}\chitil(t)\chitil(t)^\top dt - \chihT\chihTT\in\bR^{P\times P}\label{eq:R}\\
\sT &= \woT{\intoT}\chitil(t)d\Nt - \LamhT\chihT\in\bR^{P}\nonumber\\
	&= \frac{\VhT}T - \frac{M_T}{T}\chihT + \RTwo, \label{eq:s2}
}  
where the equivalent expression \eqref{eq:s2} of $\sT$ 
results from the martingale property in Lemma \ref{lem:mar}(a), 
$\VhT=\intoT\chitil(t)dM_t $ and the cross-covariance term is
\eqn{\label{eq:Rwo}
\RTwo = \woT\intoT\chitil(t)\chio(t)dt - \chihT\woT\intoT\chio(t)dt,
}
Equations \eqref{eq:R} and \eqref{eq:s2} admit a statistical interpretation. 
$R_T$ represents the empirical covariance of the candidate memory regressors. 
Provided that the stochastic integral $\VhT$ in \eqref{eq:s2} satisfies 
the conditions of Lemma \ref{lem:mar} to be a martingale, 
$s_T$ functions as a cross-covariance estimator between the truncated candidate memory regressor $\chitil(t)$ and the true memory process $\chio(t)$.

\subsection{Main Result I: The Least-Squares Estimator}
All $\Lamh,\chihT, \gT, \sT, \RT, \GT$ are fully observed, thanks to the prescribed UMCKs. 
If the empirical Gram matrix $G_T$ is positive definite, 
the LS estimator can be easily obtained using the classical matrix calculus \cite{Magn19}. 
However, two issues require attention. 
First, the empirical Gram matrix must be positive definite for finite $T$. 
Second, the closed-form LS solution requires a compact form 
for both interpretation and asymptotic analysis. 
We settle both issues in the theorem below and present the closed-form LS estimators, which first appeared in our conference version \cite{Rong26}.

\Bthm{thm:ls}{\bf Hawkes LS Estimators. }\cite{Rong26}
Denote $t_1$ the time of the first event, let $T$ be the deterministic observation time, 
and $T_0$ be the deterministic minimal affine-independence horizon as defined in \ref{A2}(*). 
Under \ref{A1} and \ref{A2}, 
if $t_1<T-T_0$, both $\RT$ and $\GT$ are positive definite, almost surely. 
Then, the LS estimator is given by
\eq{
\theh &=\sqmat{\hat\alpha\\\ch}=\argmin_{\theta\in\bR^{P+1}} J_T(\theta)
	= \GT^{-1}\gT. 
}
Further, using Schur complement \cite{Magn19}, we have 
$\GT^{-1}= \sqbra{\smat{\RT^{-1}&-\RT^{-1}\chihT\\-\chihTT\RT^{-1}&\chihTT\RT^{-1}\chihT+1}}$. Therefore, 
\eqn{
\hat\alpha &= \RT^{-1}\sT\label{eq:alph}\\
\ch &= \hat \Lam_T - \chihTT\hat\alpha.\label{eq:ch}
}
\Ethm

\pro{
Suppose $\GT>0$. By matrix perturbation \cite{Magn19}, 
it is straightforward to show that $\theh = \GT^{-1}\gT$ 
uniquely minimizes $J_T(\theta)$. 
Since the bottom-right element of $\GT$ is $1>0$, 
due to Schur complement, $\GT>0$ iff $\RT>0$. 
Then the inversion $\GT^{-1}$ using Schur complement 
directly results in the equivalent form of $\theh$ given in (\ref{eq:alph}) and (\ref{eq:ch}). 
In Appendix \ref{apxA}, 
we complete the proof by showing $\RT>0$ almost surely 
for all $T>t_1+T_0$.
}

The interpretation of the theorem is as follows. 
The LS estimators are guaranteed to exist, 
subject to a minimal data sufficiency condition.
Define $E_T= \{t_1<T-T_0\}$. 
On $E_T$, both $\RT$ and $\GT$ are positive definite almost surely, 
and the closed-form estimators \eqref{eq:alph}–\eqref{eq:ch} are well-defined. 
On the complement of $E_T$, the memory regressors are affinely dependent on $[0,T]$, 
the model is unidentifiable on that horizon, and 
$\RT$ is not positive definite. 
Consequently, $\Pr[\RT>0] = \Pr[E_T]$. 
Under \ref{A1}, $\Pr[E_T]\to1$ as $\Ttoi$. 
For finite $T$, verifying $E_T$ is then a necessary identifiability check. 
In the typical case of $T_0=0$ (see discussions after \ref{A2}), 
the condition becomes ``at least one event must be observed'', 
which is trivial in practice. 

Regarding the Gram matrix, 
prior work either assumes positive definiteness \cite{Meno18} 
or shows it only with high probability \cite{Hans15} 
(without the simple, observable check as we do). 
The equivalent estimates (\ref{eq:alph}) and (\ref{eq:ch}) 
have the following properties. 

\Brmk{\bf Mass Conservation. }
Reorganize (\ref{eq:ch}) to find
$
\NtilT = \ch T + \intoT\chitil(t)^\top dt\hat\alpha = \intoT\lamtilw(t;\theh)dt
$. 
Thus, the fitted candidate intensity integrates to the observed count,  
despite possible misspecification. 
The same property holds for EM estimators (See \cite[Result VI]{Godo20}).
\Ermk

\Brmk{\bf Relation to the Centered LS. }
The centered LS (CLS)\footnote{See \cite{Rong23} for a derivation 
for the continuous-time CLS and \cite{Solo18} for a discrete-time version.} 
sets $c=\LamhT - \chihTT\alp$ in (\ref{eq:ls}) so that 
the corresponding LS contrast $J_T'(\alp)$ only relies on $\alp$. 
The CLS separates $\alp$'s and $c$ and, therefore, offers a 
more suitable formulation for parameter constraints  \cite{Rong23,Solo18}. 
It is straightforward to see $\hat\alpha$ is the minimizer of the CLS contrast $J'_T(\alp)$ 
and $\ch$ is exactly the CLS estimator by recovering the centering. 
This shows that the (unconstrained) LS and CLS are equivalent. 
\Ermk

\mysec{Asymptotic Convergence}\label{sec:conv}
This section establishes the almost-sure asymptotic convergence of the LS estimators. 
We prove that the estimators are strongly 
consistent when the true HIR is spanned by the candidate UMCKs, 
and converge to well-defined pseudo-true parameters otherwise. 
These results strengthen our preliminary work \cite{Rong26} 
by strengthening convergence in probability to convergence w.p.1 and 
replacing the explicit ergodicity assumption 
directly on memory regressors with much more general conditions on the causal kernels. 

We begin by formally defining a correct specification. The subsequent subsections then establish the fundamental assumptions and key lemmas that underpin the asymptotic analysis. The main results are presented in Section \ref{subsec:conv}.

\Bdef{def:c}
{\bf Correct specification.}
The candidate intensity $\lamw(t;\theta)$ correctly 
specifies the true Hawkes process if \ref{A1} and \ref{A2} 
are satisfied, and there exists a unique $\alpo\in\bR^P$ such that 
$\phio(t) = \alpo^\top q(t)$, a.e. on $t\in[0,\infty)$. 
In this case, we define $\theta_0 = [\alpo^\top,c_0]^\top$ as the true parameters.
\Edef

\subsection{An Ergodic Lemma}
We present an ergodic lemma that establishes the fundamental ergodic properties required for our asymptotic analysis. 

\Blem{lem:ergsta}{\bf Ergodic Lemma.} Let $N$ be a stationary counting process satisfying \ref{A1} and 
$g_j\in\Lw\cap\Li,j\in\bN$ be deterministic functions.  
Define the stochastic processes $f_j(t)=g_j\star dN_t, j\in\bN$. 
\nlist{
\ita 
For all $n\in\bN$,  $\prod_{j\in \cP_n}f_j(t), \cP_n=\{1,2,\dotsm,n\}$ are stationary with 
\eq{
\E[\sprod{j\in \cP_n}f_j(t)]=\E[\sprod{j\in \cP_n}f_j(0)]<\infty.
} 

\itb Specially, 
\eq{
\E[\lamo(t)^n]=&\E[\lamo(0)^n]<\infty,\forall n\in\bN\\
\E[f_j(0)] =&\Lam\mat{\intoi} g_j(u)du\\
\Cov[f_1(0),f_2(0)]=&\E[f_1(0)f_2(0)]-\E[f_1(0)]\E[f_2(0)]\\
	=&\mat{\intoi\intoi} g_1(v)g_2(u)C(u-v)dvdu.
} 
\itc Further, for all $n\in\bN$, as $\Ttoi$, 
\eq{
\woT\intoT \sprod{j\in\cP_n} f_j(t)dt\to\E[\sprod{j\in\cP_n} f_j(0)], \mbox{w.p.1}.
} }
\Elem

\subsection{Kernel Conditions and the Vanishing Bias Terms}
Lemma \ref{lem:ergsta} alone is insufficient to establish the required convergence, 
as our analysis involves non-stationary truncated processes. 
We must further demonstrate that the bias terms, 
arising from unobserved pre-sample events and the martingale differences, 
vanish asymptotically w.p.1. 
To achieve this, we introduce the following regularity conditions.

\Bass{A3}{\bf Kernel regularity conditions. }\\
(a) For the true HIR, $\intoi t\phio(t)dt<\infty$.\\
(b) For the UMCKs, $\intoi tq_j(t)dt<\infty$
, $j\in\{1,\dotsm,P\}$.

\Eass

\ref{A3} is imposed to ensure that the influence of the unobserved history prior to time $0$ vanishes asymptotically in the truncated statistics \cite{Brem96}. 
This condition is mild and widely satisfied; it holds for all kernels with exponential tails (e.g., the standard Hawkes-exponential \cite{Hawk71} and Hawkes-Laguerre \cite{Godo20} models) as well as for heavy-tailed kernels with a finite mean. 

\Blem{lem:dri}{\bf Vanishing drift terms. }
Let $N$ be a stationary Hawkes process with \eqref{eq:lam}. Let $g_j\in\Lw\cap\Li,j\in\bN$ and 
$h_j\in\Lw,j\in\bN$ be deterministic functions. 
Define $\Delta f_1(t) =\intiom g_1(t-\tau)dN_\tau$ 
to be the pre-zero memory 
and $f^{A_j}_j(t) = \int_{A_j} g_j(t-\tau)dN_\tau, A_j\subseteq\bR$. 
If $\intoi t|g_1(t)|dt<\infty$, then for any $\ep>0$, as $\Ttoi$, 
the following processes converge to $0$ w.p.1: 
\eq{
&\wo{T^\ep}\intoT h_1\star \Delta f_1(t)\prod_{j=2}^n h_j\star f^{A_j}_j(t)dt,  \\
&\wo{T^\ep}\intoT h_1\star\Delta f_1(t)dN_t, \quad
\wo{T^\ep}\intoT h_1\star\Delta f_1(t)dM_t.
}
\Elem

Lemma \ref{lem:dri} will simplify the analysis of the truncated processes. 
We decompose the truncated process (e.g. $\chitil(t)=\intotm q(t-u)dN_u$) 
into the difference between the stationary process (e.g. $\chi(t)=\intitm q(t-u)dN_u$) 
and the pre-zero memory process (e.g. $\Dchi(t)=\intiom q(t-u)dN_u$). 
By setting $h_j$ in Lemma \ref{lem:dri} to the Dirac delta function $\delta$ (note that $\Lwn{\delta}=1$), the averaging terms involving the pre-zero memory process vanish w.p.1. This isolates the stationary components, allowing us to directly apply Lemma \ref{lem:ergsta}. The following Lemma \ref{lem:marint} will establish the convergence of $\sT$ as a cross-covariance estimator.

%
\Blem{lem:marint}{\bf Martingales in integral forms. }
Let $N$ be a stationary Hawkes process with \eqref{eq:lam}. Let $g_j\in\Lw\cap\Li$ be deterministic functions and 
$f_j(t)=g_j\star dN_t,\tilde f_j=g_j\star d\Ntilt$
be stochastic processes. Define  
$M^f_{j,t}=\intotm f_j(u)dM_u$ and $\tilde M^f_{j,t}=\intotm \tilde f_j(u)dM_u$.
\nlist{
\ita Both $M^f_{j,t},\tilde M^f_{j,t}$ are $\cHtm$-martingales. 
\itb 
$
\woT M^f_{j,T}\to0,\quad \woT \tilde M^f_{j,T}\to0$ w.p.1, as $\Ttoi$. 
}
\Elem

\subsection{Main Result II: Asymptotic Convergence} \label{subsec:conv}
Here, we present the key convergence results 
that establish the convergence of the LS estimators. 
We list the convergence of $\LamhT,\chihT,\RT,\RTwo,\sT$ as lemmas 
and conclude the convergence of the LS estimator by the continuous mapping theorem 
in Theorem \ref{thm:theconv} below. 

\Blem{lem:chiconv}
Under \ref{A1}-\ref{A3}, as $\Ttoi$, 
\begin{align}
\chihT &\to \mu\trieq\E[\chi(0)]=\Lam\bw_{P}, \quad\mbox{w.p.1.}, 
\end{align}
where $\bw_k$ is a $k$-dimensional vector of $1$'s. 
\Elem

\Blem{lem:Rconv}
Under \ref{A1}-\ref{A3}, 
\nlist{
\ita as $\Ttoi$, the covariance estimators
\begin{align*}
&\RT	\to \Rs, 	\quad\mbox{w.p.1.}\\
&\RTwo\to \Rswo, \quad\mbox{w.p.1.},
\end{align*}
where $\Rs,\Rswo$ exist with $\Rs>0$ and are given by
\begin{align}
\Rs	&\trieq \Var[\chi(0)] =  \E\sqbra{\chi(0)\chi(0)^\top}-\mu\muT\nonumber\\
	&= \intoi\intoi q(u)C(u-v)q(v)^\top dudv\label{eq:Rs}\\
\Rswo	& \trieq \Cov[\chi(0),\phio(0)] =\E\sqbra{\chi(0)\phio(0)}- \Gam\mu\nonumber\\
	&=\intoi\intoi q(u)C(u-v)\phio(v) dudv.\label{eq:Rswo}
\end{align}
\itb By Parseval's theorem \cite{Oppe97}, 
\begin{align}
\Rs &= \wo{2\pi}\intii  \qb(\jw)\bar C(\omega)  \qb(-\jw)^\top d\omega\label{eq:Rs}\\
\Rswo &= \wo{2\pi}\intii \qb(\jw)\bar C(\omega) \phibo(-\jw) d\omega.\label{eq:Rswo}
\end{align}
}
\Elem

\Blem{lem:sconv}
Consider $s_T$ in \eqref{eq:s2}. Under \ref{A1}-\ref{A3}, 
\nlist{
\ita $\VhT=\intoT\chitil(t)dM_t$ is an $\cH_{T-}$-martingale, 
\itb as $\Ttoi$, $\frac{\VhT}T\to0, \frac{M_T}T\to0$, w.p.1, and
\itc as $\Ttoi$, $\sT \to \Rswo$, w.p.1.
}

\Elem

The above convergences and the positive definiteness of $\Rs$ lead 
to the convergence of the LS estimators. 

\Bthm{thm:theconv}{\bf Strong consistency of the LS estimators. }
Under \ref{A1}-\ref{A3}, the LS weighting estimator $\hat\alpha$ 
and the LS background rate estimator $\ch$ both converge w.p.1 
to their pseudo-true values $\theh\to\thes=[\alps^\top,\cs]^\top$, given by 
\begin{align}
\alps &= \Rs^{-1}\Rswo\label{eq:alps}\\
\cs &=  \Lam - \muT\alps= \Lam(1-\Gams),\label{eq:cs}
\end{align}
where we define $\Gams = \bw_{P}^\top\alps$ as the {\it pseudo-true branching ratio}. 
In the case of correct specification, the LS estimators are consistent: 
$\alps = \alpo$ and $\cs = \co$.
\Ethm
\pro{
The first set of convergences follows from Lemmas \ref{lem:chiconv}-\ref{lem:sconv} and the continuous mapping theorem. 
The frequency domain expression follows 
by noticing, e.g., $\intoi\intoi q(u)C(u-v)q(v)^\top dvdu = \intoi q\star C(u)q(u)^\top du$ 
and then applying Parseval's theorem \cite{Oppe97}. 
Consistency follows by noticing that, under correct specification, 
$\phio(t)=q(t)^\top\alpo$, so we have
$\Rswo=\Rs\alpo\Ra\Rs^{-1}\Rswo=\alpo$ and $\cs=\Lam(1-\bw^\top_P\alpo) = \Lam(1-\Gam) = c_0$, by Lemma \ref{lem:1mom}.
}

Theorem \ref{thm:theconv} provides a fundamental justification for Hawkes modeling with prescribed basis kernels in two key aspects. First, under correct specification, where the true HIR lies within the span of the known basis functions, the LS estimators are strongly consistent with the true parameters. Second, since the true basis is rarely known in practice, we establish that under misspecification, the estimators still converge w.p.1 to well-defined pseudo-true parameters. This convergence is guaranteed by the existence of the limiting $\Rs$ and $\Rswo$, and the positive definiteness of $\Rs$.

Theorem \ref{thm:theconv} strengthens previous MLE results in the literature in the following aspects. 
Existing literature, e.g. \cite{Clin17} and \cite{Kwan23},  imposes explicit high-order moment assumptions on the intensity, such as $\E[\lamo(0)^{3+\ep}]<\infty$ for some $\ep>0$. In contrast, we derive these moment properties directly from the primitive conditions on the kernels. 
While analyses of MLEs often rely on weak laws of large numbers \cite{Ogat78,Clin17,Kwan23} due to the technical difficulty of handling the logarithm of the intensity, the LS framework yields convergence w.p.1. 
Another major advantage of the closed-form LS estimators is that consistency follows from the pointwise ergodic convergence. This bypasses the need for uniform laws of large numbers over the parameter space, which are strictly required for M-estimators like the MLE. 

\mysec{Central Limit Theorems}\label{sec:clt}
In this section, we derive the CLTs for the LS estimators under both 
correct specification and misspecification. 
The closed-form LS estimators allow us to develop the 
asymptotic Gaussian covariances with explicit, interpretable structures.

The derivation under correct specification, as we shall see below, 
follows almost directly from the functional martingale CLT. However, 
the misspecified case presents a unique challenge 
as it contains both a martingale and a non-stationary bias component 
that also contributes to the asymptotic covariance. 
We resolve this by extracting the martingale component from the bias term, allowing us to characterize the joint distribution via a unified martingale framework. 
We note that these results could alternatively be derived using the more abstract mixing properties of the process \cite{Brad07}. However, given the inherent martingale structure of the estimator error, the martingale CLT approach provides a more direct path to explicit covariance expressions.

\subsection{The Martingale Representation Motivation} 
Let $\MtilxiT=\intoT\xitil(t)dM_t$. 
Using the $\sT$ expression \eqref{eq:s2}, the LS estimators \eqref{eq:alph},\eqref{eq:ch} and the pseudo-true values \eqref{eq:alps},\eqref{eq:cs}, we can express the joint parameter error as
\eqn{
	\sqrt{T}(\theh-\thes)
=	G_T^{-1}\wo{\sqrt{T}}\MtilxiT+B_T+o_p(\bw_{P+1}),\label{eq:dthe}
}
where $G_T^{-1}=\sqbra{\smat{\RT^{-1}&-\RT^{-1}\chihT\\-\chihTT\RT^{-1}&\chihTT\RT^{-1}\chihT+1}}$ is exactly the block inverse of the empirical Gram matrix via Schur complement, as in Theorem \ref{thm:ls},  
and $B_T = \sqbra{\smat{\RT^{-1}\BaT\\\BcT-\chihTT\RT^{-1}\BaT}}$ 
is the bias term, where 
$\BaT = \sqrt{T}(\RTwo-\RT\Rs^{-1}\Rswo)$ represents the bias introduced by the truncation of the memory process and the possible misspecification, 
and $\BcT=\sqrt{T}(\LamhT-\Lam)-\sqrt{T}(\chihT-\mu)^\top\alps$ 
represents the variation in the empirical rate.

Under correct specification and \ref{A1}-\ref{A3}, 
the bias terms 
\eq{
\BaT=	&\sqrt{T}(\RTwo-\RT\Rs^{-1}\Rswo) 
= 	\wo{\sqrt{T}}(\RTwo - \RT\alpo)\\
=	&\wo{\sqrt{T}}{\intoT\chitil(t)\Dchio(t)dt - \wo{\sqrt{T}}\chihT\intoT\Dchio(t)dt}\to0, 
}
w.p.1, thanks to Lemma \ref{lem:chiconv} and Lemma \ref{lem:dri}, 
and 
\eq{
	\BcT
=	&\sqrt{T}(\NtilT - \woT\alpo^\top\mat{\intoT}\chitil(t)dt - (\Lam-\Lam\bw^\top\alpo))\\
=	&\frac{M_T}{\sqrt{T}} + \wo{\sqrt{T}}\intoT\Dchio(t)dt=\frac{M_T}{\sqrt{T}} + o_p(1), 
}
thanks to Lemma \ref{lem:dri}. 
Thus, under correct specification, $\sqT(\theh-\thes)=\wo{\sqT}\MtilxiT+o_p(\bbw{P+1})$ admits a scaled martingale representation. 
However, under misspecification, 
the bias term $B_T$ will be shown to persist and contribute to the asymptotic covariance under misspecification.

Define $\xi(t) = [\chi(t)^\top,1]^\top$ as the full extended regressor.
Clearly, by previous lemmas, 
$\GT\to\Gs=\Gs^\top\trieq\E[\sqbra{\smat{\chi(0)\\1}}\sqbra{\smat{\chi(0)\\1}}^\top]=\E[\xi(0)\xi(0)^\top]=\sqbra{\smat{\Rs+\mu\mu^\top&\mu\\\mu^\top&1}}$ w.p.1 as $\Ttoi$, 
and by the continuous mapping theorem, 
$G_T^{-1}\to \Gs^{-1}=\sqbra{\smat{\Rs^{-1}&-\Rs^{-1}\mu\\-\muT\Rs^{-1}&\muT\Rs^{-1}\mu+1}}$.

Now the remaining tasks are clear. 
In the next subsection, we apply a functional martingale CLT to $\wo{\sqT}\MtilxiT$ 
and check the required conditions to develop the CLT under correct specification. 
In the last subsection, we address model misspecification by deriving martingale representations for $\BaT$ and $\BcT$ and jointly applying the functional CLT.

\subsection{Main Result III: CLT under Correct Specification}
We will apply the following functional CLT for martingales, 
which translates the classic \cite[Chapter VIII, 3.24]{Jaco13} into point process notations. 

\Blem{lem:mclt}{\bf Functional CLT for martingales. }\cite[Chapter VIII, 3.24]{Jaco13}
Let $f_T(t)$ be a vector $\cHtm$-predictable process. 
Define the process $X_\tau^T = \intotT f_T(t)dM_t, \tau\in[0,1]$. 
If 
\nlist{
\ita $\E[\intotT \Ver{f_T(t)}^2\lamo(t)dt]<\infty$, $\forall \tau\in[0,1], T>0$,\footnote{Given the structure of $X_\tau^T$, condition (a) is equivalent to saying $F(\tau)=X_\tau^T$ is a square-integratable $\cH_{\tau T-}$-martingale for any fixed $T$.}
\itb $\ang{X^T}_\tau\convp \tau\Sig$, as $\Ttoi$, for all $\tau\in[0,1]$, 
\itc (Lindeberg) for any $\ep>0$,\\ 
$\intotT\Ver{f_{T}(u)}^2\cI_{\Ver{f_{T}(u)}\geq\ep}\lamo(u)du\convp0$ as $\Ttoi$,
} 
then $X^T_\tau\convd \Sig^{\haf}W_\tau,\tau\in[0,1]$, where 
$W_\tau$ is a standard Brownian motion, and the convergence in distribution is in the Skorokhod space $\cD[0,1]$. 
\Elem

Checking the above conditions yields the CLT under correct specification. 

\Bthm{thm:cltc}{\bf CLT under correct specification. }
Under \ref{A1}-\ref{A3} and correct specification, 
\eq{
&\sqrt{T}(\theh-\theta_0)\RA\cN(0,\Sigoth)\\
\mbox{where, }\quad\Sigoth
	 	&= \Gs^{-1}\E\sqbra{\lamo(0)\xi(0)\xi(0)^\top}\Gs^{-1}\\
		&=\sqbra{\smat{\Sigoalp & -\Sigoalp\mu+\alpo\\ -\muT\Sigoalp+\alpo^\top&\sigoc}}\\
\Sigoalp 	&= \Rs^{-1}\Sigo\Rs^{-1}>0\\
\sigoc	&
		= \muT\Sigoalp\mu + \Lam(1-2\Gam)>0\\
\Sigo		&= \E[\lamo(0)(\chi(0)-\mu)(\chi(0)-\mu)^\top]>0.
}
\Ethm
\pro{
Split $\wo{\sqrt{T}}\intoT\xitil(t)dM_t = \wo{\sqrt{T}} \intoT\xi(t)dM_t - \wo{\sqrt{T}}\intoT [\Dchi(t)^\top, 0]^\top dM_t$. The second term vanishes 
because each $\wo{\sqrt{T}}\intoT\Dchij(t)dM_t\to0$ w.p.1 as $\Ttoi$ by Lemma \ref{lem:dri}. 
Now we check the conditions in Lemma \ref{lem:mclt} for the first term. Let 
$X_\tau^T = \wo{\sqrt{T}}\intotT\xi(t)dM_t$. 
Condition (a) is satisfied because $\woT\E[\intotT\Ver{\xi(t)}^2\lamo(t)dt]=\tau \E[\Ver{\xi(0)}^2\lamo(0)]$ by Lemma \ref{lem:ergsta}. 
Then, $X_\tau^T$ has a quadratic variation 
$\ang{X^T}_\tau=\wo{T}\intotT\xi(t)\xi(t)^\top \lamo(t)dt\to\tau \E[\lamo(0)\xi(0)\xi(0)^\top]$ w.p.1, as $\Ttoi$ by Lemma \ref{lem:ergsta}. 

We check the Lindeberg condition (c). 
By Markov's inequality, we only need to show that the following expectation tends to $0$. Use $\cI_{\Ver{\cdot}\geq\ep}\leq \frac{\Ver{\cdot}}\ep$ to find
\eq{
	&\mat{\E[\intotT\Ver{\xi(t)/\sqrt{T}}^2\cI_{\Ver{\xi(t)/\sqrt{T}}\geq\ep}\lamo(t)dt]}\\
\leq	&\mat{\wo\ep\E[\intotT\Ver{\xi(t)/\sqrt{T}}^3\lamo(t)dt]}\\
=	&\mat{\wo{\ep T^{\frac32}}\intotT\E[\Ver{\xi(t)}^3\lamo(t)]dt.}
}
Use H\"older's inequality on the expectation and Jensen's inequality on the norm to find 
\eq{
	&\E[\Ver{\xi(t)}^3\lamo(t)] \leq \E[\Ver{\xi(t)}^4]^{3/4}\E[\lamo(t)^4]^{1/4}\\
\leq 	&((P+1)\E[1+\sjwP\chij(t)^4])^{3/4}\E[\lamo(t)^4]^{1/4}\\
=	&(P+1)^{\frac34}({1+\sjwP\E[\chij(0)^4]})^{\frac34}\E[\lamo(0)^4]^{\frac14}<\infty.
} 
We thus have $\E[\intotT\Ver{\xi(t)/\sqrt{T}}^2\cI_{\Ver{\xi(t)/\sqrt{T}}\geq\ep}\lamo(t)dt]=O(\wo{\sqrt{T}})$, for any $\tau\in[0,1]$. 
By Lemma \ref{lem:mclt}, we conclude $\wo{\sqrt{T}}\intoT\xitil(t)dM_t = X_1^T\RA\cN(0,\E[\lamo(0)\xi(0)\xi(0)^\top])$. 
Then, by Slustky's theorem \cite{Van00} and straightforward matrix calculations, the results follow. 
}

Theorem \ref{thm:cltc} establishes that the LS estimator for the Hawkes process is asymptotically normal, 
converging to the true parameters at the standard $\sqrt{T}$ rate. 
The block partition of the asymptotic covariance $\Sigma_0^\theta$ offers a clear geometric interpretation. It explicitly separates the uncertainty associated with the weighting parameters $\alpo$ from that of the background rate $c_0$, while capturing their dependence through the cross-covariance terms.

To assess statistical efficiency, we compare this result to the theoretical lower bound. The Cram\'er-Rao Bound (CRB) for the parameters of a stationary Hawkes process is the inverse of the Fisher Information Matrix given by  \cite{Godo20} 
\eq{
\Sigcrb=\E[\wo{\lamo(0)}\xi(0)\xi(0)^\top]^{-1}.
}
Using Jensen's inequality and the matrix \CSi\ \cite{Magn19}, 
it is straightforward to find that\footnote{See Supplimentary Material.}
$\Sigcrb\leq\Sigoth$, suggesting that the LS estimator is not optimal, 
compared to the MLEs \cite{Clin17,Ogat78}. 
The identified efficiency gap motivates a weighted LS identification, which we hope to tackle in the future. 

\subsection{Main Result IV: CLT under Misspecification} 
Here, we derive the asymptotic normality of the misspecified estimator by establishing the martingale representation of the bias term $B_T$ and applying Lemma \ref{lem:mclt}. Central to this derivation is the expression of the asymptotic covariance via the Hawkes resolvent $\kap$, which emerges from the martingale representation of the intensity deviation. Using this key lemma, we identify the martingale representations of $\BcT$ and $\BaT$ and conclude the proof using the functional martingale CLT.

We define the centered measure $\nu(dt) = dN_t-\Lam dt$,  
$\eta(t) = \intiom \phio(t-u)\nu(du)=\Dchio(t)-\Lam\intti\phio(u)du$, 
and $\zeta(t) = \sum_{n=0}^\infty \phio^{\star n}(t) = \delta(t)+\kap(t)$,
whose LT is $\bar\zeta(s) = \wo{1-\bar\phi_0(s)}$. 
Clearly, 
$
\Lwn{\zeta} = 1 + \Lwn{\kap} = \wo{1-\Gam}$. 
We define the (deterministic) \pt\ Hawkes HIR $\phis(t)=\alps^\top q(t)$ and 
the (deterministic) Hawkes HIR difference $\Dphi(t) = \phio(t)-\phis(t)$. 
Clearly both $\phis\in L^p[0,\infty)$ and $\Dphi\in L^p[0,\infty)$ for any $1\leq p\leq\infty$ and under \ref{A3}, 
\eqn{
\intoi t|\phis(t)|dt<\infty, \quad \intoi t|\Dphi(t)|dt<\infty. \label{eq:tphi}
}
We will write $g\star dM_t = \intot g(t-u)dM_u$ for a vector function $g:[0,\infty)\mapsto\bR^k$ and $|g_j|\star dM_t=\intot |g_j(t-u)|dM_u$ for a scalar function $g_j: [0,\infty)\mapsto\bR$.

\Blem{lem:lamm}{\bf Martingale representation of $\lamo(t)-\Lam$. }Under \ref{A1}, 
$
\lamo(t)-\Lam = \kap\star dM_t + \zeta\star\eta(t)$. 
Consequently, for a measurable vector function $g:[0,\infty)\mapsto\bR^k$, 
$
\intot g(t-u)\nu(du) 
=g\star\zeta\star dM_t + g\star\zeta\star\eta(t)$. 
\Elem
\pro{
By expanding the intensity and using the relation $\Lam=\frac{c_0}{1-\Gam}$ (Lemma \ref{lem:1mom}), we find $(\lam(t)-\Lam)-\intot\phio(t-u)(\lam(u)-\Lam)du = \phio\star dM_u+\eta(t)$. Since $\zeta(t) = \sum_{n=0}^\infty \phio^{\star n}(t)$ 
is the convolutional inverse of $\delta(t) - \phio(t)$, 
we solve the Volterra equation \cite{Brem96} to get 
$\lam(t)-\Lam = (\zeta\star\phio)\star dM_t+ \zeta\star\eta(t) = \kap\star dM_t+\zeta\star\eta(t)$, as quoted. 
}

If we can rewrite the bias terms $\BaT,\BcT$ in forms of 
the centered measures, we can subsequently replace these measures with martingale measures using the lemma above. The terms involving the pre-zero memory $\eta$ vanish by Lemma \ref{lem:dri}, isolating a martingale core that facilitates the application of the functional martingale CLT.

\Blem{lem:bct}{\bf Martingale representation of $\BcT$. }Under \ref{A1}-\ref{A3}, 
$
\BcT = \wo{\sqrt{T}}\frac{1-\Gams}{1-\Gam}M_T+o_p(1)$.
\Elem

\Blem{lem:bat}{\bf Martingale representation of $\BaT$. }Under \ref{A1}-\ref{A3}, 
$\BaT = \wo{\sqrt{T}}\intoT(\chitilha(t)-\muha)dM_t+o_p(\bbw P)$, where 
$\chitilha(t) = \ha\star d\Ntilt$ with $\Lam \intoi\ha(u)du\trieq\muha$, and 
$h^\alp(u)=[\ha_1(u),\dotsm,\ha_P(u)]^\top=\Wtil(u)-\Wtil\star \phio(u)\in\bR^P$, 
where $\Wtil(u) =	\cI_{u\geq0}\intoi a(t+u)b(t)dt + a(t)b(t+u)dt$, with  
$a(u)=q\star\zeta(u)\in\bR_{\geq0}^P, b(u)=\Dphi\star\zeta(u)\in\bR$. 
We have $\ha_j\in\Lw\cap\Li$, 
$\intoi u|\ha_j(u)|du<\infty$.

\Elem

Lemma \ref{lem:bat} presents a significant challenge 
for which we provide a four-stage proof in Appendix \ref{apxC}. 
Define $h(t) = q(t)+\ha(t), \chitilh(t) =h\star d\Ntilt, 
\chihh(t)=h\star dN_t, 
\muh = \intoi h(t)dt = \mu + \muha$, 
$\xitilh(t) = \xitil(t) + [\chitilh(t)^\top, 0]^\top=[\chitil(t)^\top+\chitilh(t)^\top,1]^\top$,  
and $\kappa = \frac{1-\Gams}{1-\Gam}$. 
Let $\MtilxihT=\intoT\xitilh(t)dM_t$. 
The scaled estimation error, ignoring the $o_p(\bbw{P+1})$ terms, is given by
\eq{
	&\sqT(\theh-\thes)
=	\sqbra{\smat{\RT^{-1} & -\RT^{-1}(\chihT+\muha)\\ -\chihT^\top\RT^{-1} & \chihT^\top\RT^{-1}(\chihT+\muha) +\kappa}}\frac{\MtilxihT}{\sqT}. 
}

\Bthm{thm:cltmis}{\bf CLT under misspecification. }
Under \ref{A1}-\ref{A3},  
$\sqrt{T}(\theh-\thes)\RA\cN(0,\Sigsth)$,
where
\eq{\quad\Sigsth  	=& H\E\sqbra{\lamo(0)\xi_h(0)\xi_h(0)^\top}H^\top\\
		=&\sqbra{\smat{\Sigsalp & -\Sigsalp\mu+\kappa\alp_h\\ -\muT\Sigsalp+\kappa\alp_h^\top&\sigsc}}\\
\Sigsalp 	=& \Rs^{-1}\Sigs\Rs^{-1}\\
\sigsc	
		=& \muT\Sigsalp\mu + \Lam(\kappa^2-2\kappa\alp_h^\top\bbw P)\\
\Sigs		=& \E[\lamo(0)(\chihh(0)-\muh)(\chihh(0)-\muh)^\top]\\
\alp_h	=& \Rs^{-1}\Rsho\\
\Rsho	=& \Cov[\chihh(0), \chio(0)]\\
		=&\wo{2\pi}\intii \bar h(\jw)\phibo(-\jw)\bar C(\omega)d\omega\\
H		=& \sqbra{\smat{\Rs^{-1} & -\Rs^{-1}\muh\\ -\muh^\top\Rs^{-1} & \mu^\top\Rs^{-1}\muh+\kappa}}.
}
\Ethm
\pro{
Let $h_j(u)$ be the $j$-th element of $h(u)$. 
Since $q_j, \ha_j\in\Lw\cap\Li\Ra h_j\in\Lw\cap\Li$ and $\intoi u|h_j(u)|du\leq\intoi u|q_j(u)|du+\intoi u|\ha_j(u)|du<\infty$, 
the proof proceeds analogously to that of Theorem \ref{thm:cltc}.
}

To the best of our knowledge, Theorem \ref{thm:cltmis} provides the first explicit analytical derivation of the asymptotic covariance matrix for any Hawkes estimators under model misspecification, while also serving as the first rigorous justification for the asymptotic normality of Hawkes LS estimators. 
By characterizing the variance components via spectral integrals, the theorem reveals that the asymptotic covariance is structured around the uncertainties associated with the weighting parameters $\alpha_*$, the background rate $c_*$, and their cross-covariance. 

Under correct specification, where $h(t)=q(t)$, $\kappa=1$, and $\Rsho=\Rswo=\Rs\alpo$, Theorem \ref{thm:cltmis} coincides with Theorem \ref{thm:cltc}. 
Under possible misspecification, the explicit formula for the asymptotic covariance $\Sigsth$ facilitates robustness analysis. For instance, given a family of true Hawkes HIRs (e.g., exponential) and a prescribed UMCKs (e.g., Hawkes-Laguerre), $\Sigsth$ can be calculated explicitly using the spectral integrals and stochastic sampling. 
While $\Sigsth$ is guaranteed to be positive semidefinite, verifying its positive definiteness requires a case-by-case analysis based on these explicit calculations.

Furthermore, this established asymptotic normality lays the groundwork for Generalized Method of Moments (GMM) tests \cite{Hall05} for model comparison, offering a potential alternative to current likelihood-ratio frameworks. Such GMM tests would benefit from the closed-form nature of LS estimators with a potential for online implementation, and from the availability of CLTs under misspecification to conduct comparisons of non-overlapping models where at least one model is inevitably misspecified \cite{Vuon89,Rive02}.

\mysec{Numerical Study}\label{sec:sim}
In this section, we present simulation examples to illustrate the performance of the LS estimators and verify their asymptotic properties 
under both correct model specification and misspecification. 
We will first demonstrate the numerical calculations of 
the explicit \pt\ parameters 
and the asymptotic covariances for an asymptotic robust analysis, 
and then run the LS identification to compare the empirical error means and covariances against their true values.

\subsection{An asymptotic robustness analysis}
We run an asymptotic robustness analysis of the Hawkes-Laguerre approximation \cite{Godo20,Ogat82} for exponential HIRs. The \HL\ model employs an Erlang basis, which is a linear transformation of the orthonormal Laguerre basis of the same order \cite{Wahl91}. Our analysis reveals that while the Erlang basis exhibits strong robustness concerning its first-order statistics, its estimation variance explodes as the model order increases, 
whose underlying cause we explicitly identify.

We consider the true intensity 
$\lamo(t) = \co + \intitm\phio(t-u)dN_u$,
where the true HIR $\phio(t-u)=\alpo^\top p(t), p(t)=[p_1(t),\dotsm,p_K(t)]^\top$.  
We consider the true UMCKs 
$p_k(t) = \beta_ke^{-\beta_k t}$, 
whose LT are $\bar p_k(s) = \frac{\beta_k}{s+\beta_k}$.  
We set $c_0=1, K=3, \alpo = [0.3, 0.2, 0.2]^\top, \beta_0 = [\beta_1,\beta_2,\beta_3]^\top = [2,6,16]^\top$, 
We will use the prescribed UMCKs $q(t)=[q_1(t),\dotsm,q_P(t)]^\top$ to approximate the true HIR. 
Under misspecification, we consider the Hawkes-Laguerre basis 
$q(t) = [q_1(t),\dotsm,q_{P}(t)]^\top$ with 
$q_j(t) = \frac{\rho^j t^{j-1}}{(j-1)!}e^{-\rho t}$, 
whose LT is $\bar q_j(s) = (\frac{\rho}{s+\rho})^j$. 
\ref{A1}-\ref{A3} are satisfied. 

Because the \pt\ parameters $\alps$, the asymptotic covariances $\Sigoth, \Sigsth$, and the UMCK LTs possess explicit spectral forms, they are readily evaluated in the frequency domain. The Supplementary Material details our computational approach where we use numerical integration in the frequency domain to evaluate $\Rs,\Rswo,\muha,\Wtil(t)$, and $\ha(t)$, and employ Monte Carlo simulations to estimate the third-order moments $\E[\lamo(0)\xi(0)\xi(0)^\top]$ and $\E[\lamo(0)\xi_h(0)\xi_h(0)^\top]$, 
and, thereby, obtain $\alps, \Sigoth$, and $\Sigsth$ for the robustness analysis.

We set the \HL\ exponent $\rho=5$. 
We vary the \HL\ order $P\in\{1,2,3,4,5\}$. 
We find 
the \pt\ parameters 
\eq{
\mat{\alps&=0.67,&\sqbra{\smat{0.74\\-0.10}},&\sqbra{\smat{0.89\\-0.62\\0.43}}, &\sqbra{\smat{0.92\\ -0.77\\0.68\\-0.15}},&\sqbra{\smat{0.96\\-1.04\\1.44\\-1.06\\0.41}},\\
\Gams&=0.67, &0.64, &0.70, &0.69, &0.70,\\
c_* &=1.10,&1.19,&1.01,&1.04,&1.00,}
} 
for the increasing values of $P$, respecrtively.

We find that the pseudo-true background rate and the pseudo-true branching ratio 
are close to their true values $\Gam=0.7$ and $c_0=1$, 
and when $P=5$, the \pt\ values agree with the true values up to numerical error. 
$\alps$ contains negative entries. However, the positivity of $\alps$ is only a sufficient condition 
for the required condition $\phis(t)>0,\forall t>0$. 
We numerically verified 
$\min_{t>0}\phis(t)>0$ for all $P$, 
so the candidate intensities are valid at the pseudo-true parameters.  

\begin{figure}[t]
\hfill
\begin{minipage}[t]{1\linewidth}
\centering{\includegraphics[width=8.5cm]{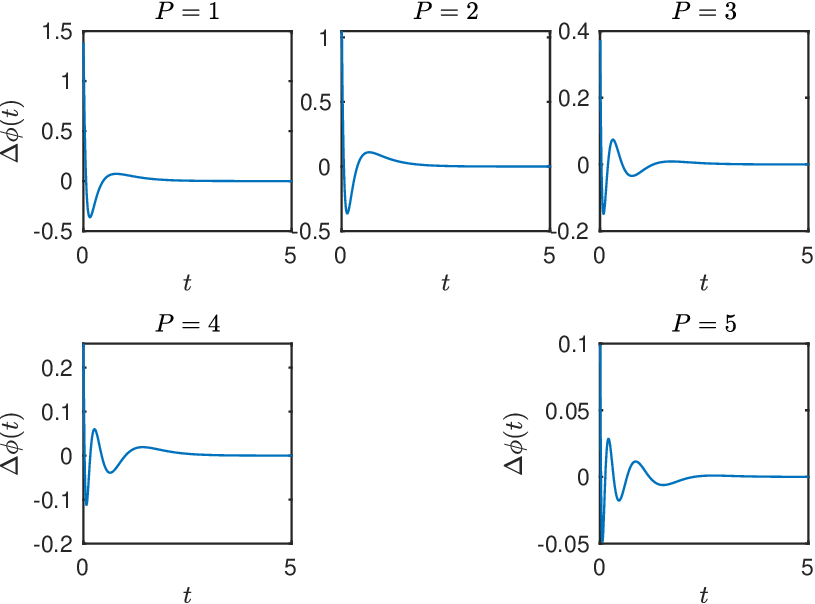}}
\end{minipage}
\caption{Plots of the HIR estimation error $\Dphi(t)$: $\Dphi(t)$ shrinks as the \HL\ model order $P$ increases 
with the relative errors $\frac{\intoi\Dphi(t)^2dt}{\intoi\phio(t)^2dt}\times100\%=5.1\%, 3.58\%, 0.40\%, 0.23\%, 0.033\%$, under $P=1,2,3,4,5$, respectively.}
\label{fig:Err}
\end{figure}

We compute the $L^2$ HIR error $\intoi \Dphi(t)^2dt=\wo{2\pi}\intii |\Dphib(\jw)|^2d\ome$ in the frequency domain. The resulting relative errors $\frac{\int_0^\infty \Delta\phi(t)^2dt}{\intoi\phio(t)^2dt} \times 100\%$ are $5.1\%, 3.58\%, 0.40\%, 0.23\%, 0.033\%$ for increasing values of $P$, respectively. We plot the corresponding HIR differences $\Delta\phi$ in Fig. \ref{fig:Err}. As the Hawkes-Laguerre order $P$ increases, the approximation quality of the HIR improves significantly, confirming the robustness of the Hawkes-Laguerre model in the first-order statistics.

To evaluate the asymptotic error covariances under both specification regimes, we simulate $L=3000$ trajectories using the thinning algorithm \cite{Ogat82} with an observation period $T=3200$ to sample the asymptotic covariances. Under correct specification, we find 
$\Sigoth = \sqbra{\smat{
   3.59   &-4.22    &1.37   &-2.29\\
   -4.22    &6.85   &-2.89    &1.20\\
    1.37   &-2.89    &1.64   &-0.24\\
   -2.29    &1.20   &-0.24    &5.65}}$ with the Frobenius norm 
   $\Ver{\Sigoth}_F=12.8$. 
We also sample the CRB in the frequency domain to find 
$\Sig_{CRB} = \sqbra{\smat{
   2.50   &-2.99    &1.02   &-1.43\\
   -2.99    &4.78   &-2.08    &1.07\\
    1.02   &-2.08    &1.22   &-0.30\\
   -1.43    &1.07   &-0.30   & 3.15}}$. 
$\Sigoth-\Sig_{CRB}>0$ with eigenvalues $3.50,0.005,0.20,2.39$ 
 and the Frobenius norm $\Ver{\Sigoth-\Sig_{CRB}}_F=4.24$. 
This indicates that under correct model specification, the asymptotic covariance of the LS estimation error closely approximates the CRB.

\begin{figure}[t]
\hfill
\begin{minipage}[t]{1\linewidth}
\centering{\includegraphics[width=8.5cm]{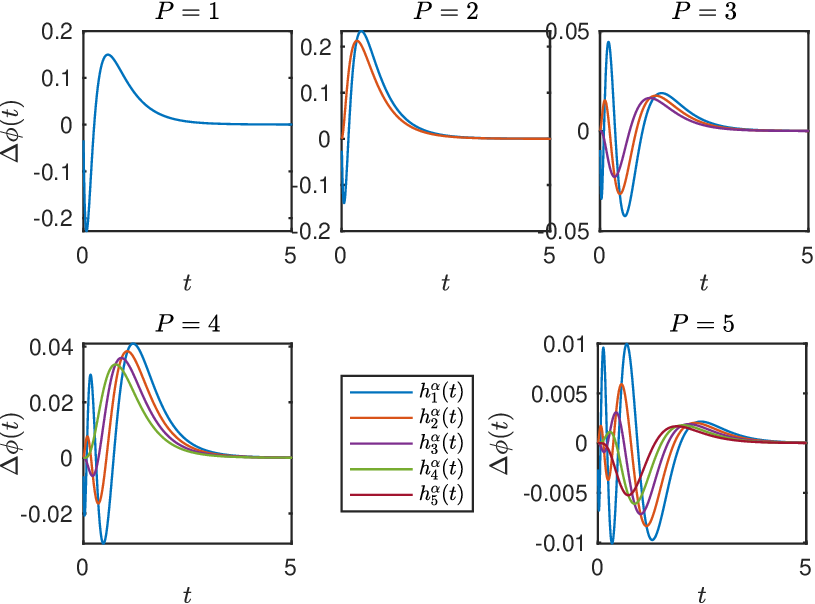}}
\end{minipage}
\caption{Plots of $\ha(t)$: $\ha(t)$ shrinks as $P$ increases, 
suggesting that $\E[\lamo(0)\xi_h(0)\xi_h(0)^\top]$ stays close to 
its correctly specified counterpart $\E[\lamo(0)\xi(0)\xi(0)^\top]$.}
\label{fig:h}
\end{figure}

 Under misspecification, we first plot $\ha(t)$ in Fig. \ref{fig:h}. 
 We find 
$\muha = 0.35, 0.63\bbw2, 0.033\bbw3, 0.129\bbw4, -0.0046\bbw5$ 
and the Frobenius norms of the asymptotic covariances $\Sigsth$ 
are $\Ver{\Sigsth}_F=4.49, 7.85, 19.53, 125.05, 938.27$,  
for the increasing values of $P$, respectively. 
This reveals a clear contradiction. On one hand, as $P$ increases, the shrinking magnitudes of $|\ha(t)|$ and $|\muha|$ suggest that $\E[\lamo(0)\xi_h(0)\xi_h(0)^\top]$ remains close to its correctly specified counterpart, $\E[\lamo(0)\xi(0)\xi(0)^\top]$. On the other hand, the error variance explodes. This inflation is driven by the scaling matrix $H$. Specifically, while Lemma \ref{lem:Rconv}(a) guarantees $\Rs > 0$, the condition number of $\Rs$ degrades with $P$: the smallest eigenvalue approaches $0$ ($0.007$ when $P=5$) while the largest grows ($71.9$ when $P=5$). Consequently, while the \HL\ model's mean is robust, its variance scales poorly. This instability could be resolved by adopting the true orthonormal Laguerre basis \cite{Wahl91} instead of the simplified Erlang basis, which would strictly bound the eigenvalues of $\Rs$. Since the true Laguerre basis is not nonnegative (violating \ref{A2}), we reserve its robustness analysis for future work.

 \begin{figure}[t]
\hfill
\begin{minipage}[t]{1\linewidth}
\centering{\includegraphics[width=8.5cm]{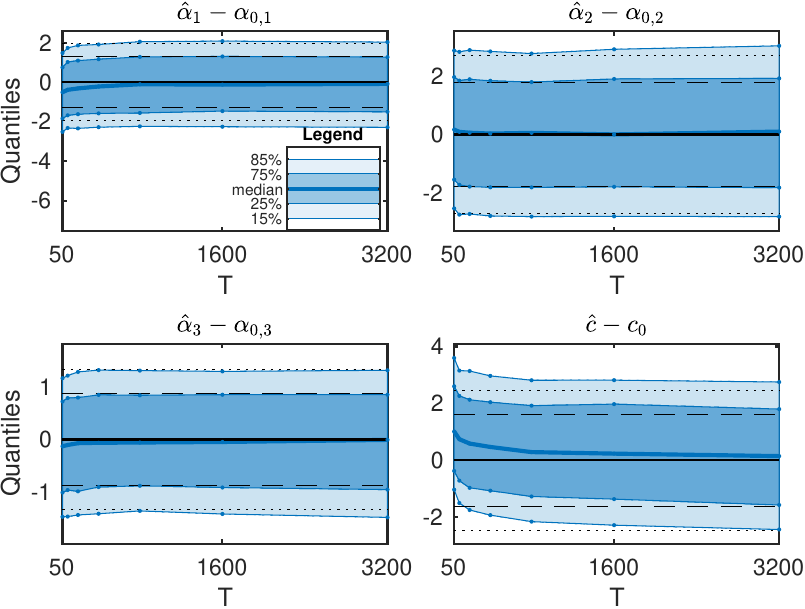}}
\end{minipage}
\caption{Quantiles of the scaled estimation error $\sqrt{T}(\theh-\theta_0)$ under correct model specification. The empirical quantiles of the LS estimates (blue) closely match the theoretical asymptotic quantiles (black).}
\label{fig:Ec}
\end{figure}

\begin{figure}[t]
\hfill
\begin{minipage}[t]{1\linewidth}
\centering{\includegraphics[width=8.5cm]{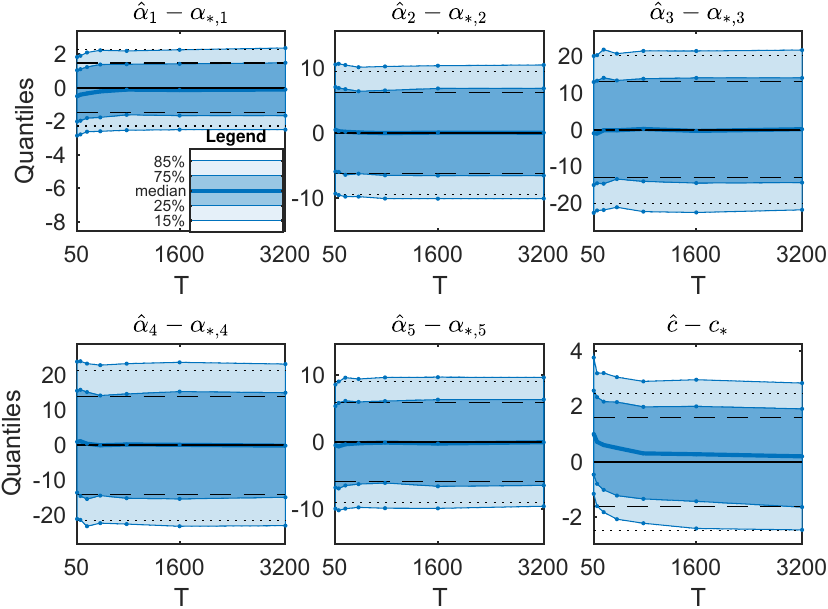}}
\end{minipage}
\caption{Quantiles of the scaled estimation error $\sqrt{T}(\theh-\thes)$ under model misspecification. The empirical quantiles of the LS estimates (blue) closely match the theoretical asymptotic quantiles (black). The error variance under misspecification is much higher than that under correct specification. }
\label{fig:Emis}
\end{figure}

%
\subsec{LS Fitting}
We now evaluate the LS identification approach under both correct model specification and misspecification, demonstrating that the empirical results closely align with the theoretical asymptotic properties established in our theorems. 
The simulation model setup is identical to 
the previous subsection, however, we consider a fixed \HL\ model order $P=5$ 
and a varying observation period $T=50,100,200\dotsm,3200$. 
We simulate $L=3000$ trajectories at each $T$ and 
fit the data using the LS identification \eqref{eq:alph}, \eqref{eq:ch}. 
A recursive calculation of $\RT,\sT$ is used 
for numerical efficiency for \HL\ basis \cite{Godo20}; we omit the computational details here.

Fig. \ref{fig:Ec} plots the $15\%$, $25\%$, $50\%$, $75\%$, and $85\%$ empirical quantiles of the scaled estimation error $\sqT(\theh-\theta_0)$ against their theoretical counterparts under correct model specification. The observed zero median aligns with the expected asymptotic consistency, and the close agreement between the empirical and theoretical quantiles validates the established CLT. Under misspecification, Fig. \ref{fig:Emis} displays the corresponding quantiles for $\sqT(\theh-\thes)$, utilizing the $P=5$-th order \HL\ model with the pseudo-true parameter $\thes$ obtained in the previous subsection. Here, we observe a similarly strong alignment with the theoretical asymptotic distribution. 
Furthermore, we observe that under misspecification, the scaled estimation error exhibits a significantly larger asymptotic covariance than in the correctly specified case, which corroborates the robustness analysis presented in the previous subsection.

\mysec{Conclusions and Future Work}\label{sec:con}
This paper has justified a continuous-time least-squares identification framework for Hawkes processes with prescribed unit-mass causal kernels. Under a mild finite-horizon affine-independence condition, we proved that the empirical Gram matrix is almost surely positive definite beyond a simple data-sufficiency condition, guaranteeing the existence and uniqueness of closed-form LS estimators. Under general kernel moment conditions, we established strong consistency: the estimators converge almost surely to the true parameters under correct specification and to explicit pseudo-true parameters under misspecification, characterised via spectral integrals of the Hawkes process.

Building on a martingale decomposition of the estimation error, we derived central limit theorems for the LS estimators in both regimes. Under correct specification, the asymptotic covariance separates the contributions of the impulse-response weights and the background rate and can be compared directly with the Cram\'er–Rao bound. Under misspecification, we obtained an explicit expression for the asymptotic covariance. Numerical studies confirm that the empirical distributions of the scaled LS errors agree closely with the theoretical Gaussian limits, and illustrate a robust trade-off: Erlang bases can approximate exponential kernels very well in mean, while for larger basis orders the eigenvalues of the Gram matrix, while still positive, become increasingly disparate, leading to substantial growth in the estimation variance. 

Several directions remain for future work. Extending the theory to multivariate Hawkes processes, with structured regularisation and sparsity, is of practical interest. The robustness analysis suggests that alternative bases such as orthonormal Laguerre functions could improve numerical conditioning, but require relaxing the non-negativity conditions imposed here. Improving statistical efficiency via weighted LS schemes and weakening the kernel moment assumptions to cover heavier tails are natural theoretical extensions. Finally, the explicit pseudo-true parameters and asymptotic covariances provide the key ingredients for Generalised Method of Moments model-comparison tests and for incorporating identification error into the design and analysis of event-triggered control and event-based sensing systems.

\renewcommand{\thesection}{\Alph{section}}
\setcounter{section}{0}
\noleminapx
\myapp{Proofs for Section \ref{sec:id}} \label{apxA}
{\it Proof of Theorem \ref{thm:ls}.} 
As discussed before, we are left to show $\RT>0$. Set $\gam(t) = x^\top\chitil(t)$. For any $x\neq0\in\bR^{P}$, we have
$x^\top\RT x = \woT\intoT \gam(t)^2dt - (\woT\intoT \gam(t)dt)^2\geq  0$, 
by the Cauchy-Schwarz inequality, with equality iff 
$f$ is a constant a.e. on $[0,T]$. 
By definition, $\chitil(t) = \intotm q(t-u)d\Nu=\sum_{t_r\in(0,t)} q(t-t_r)$. 
So for any $T>t_1+T_0$, $x^\top\chitil(t)=d$, for almost all $t\in[t_1, \min\{t_1+T_0, t_2\}]$ iff 
$x^\top q(t)=d$, for almost all $t\in[0,\min\{T_0,t_2-t_1\}]\subset[0,T_0]$, which 
contradicts \ref{A2}(b). 
Therefore, the inequality is strict and $\RT>0$, w.p.1 given $t_1<T-T_0$. 
\hfill$\square$

\myapp{Proofs for Section \ref{sec:conv}}\label{apxB}
{\it Proof of Lemma \ref{lem:ergsta}. }
{\it Stationarity:} By the generalized Campbell's theorem \cite[Chapter 6.2]{Dale03}, 
the stationarity follows directly 
since the counting process $N_t$ is stationary 
and the integrals all start from $-\infty$, guaranteeing time-invariant moments.

{\it Finite moments: } 
Denote $[\tau]_1^{k}=[\tau_1,\dotsm,\tau_{k}]^\top$. 
From Lemma \ref{lem:hmom}, we have  
$\int_{[\tau]_1^{n-1}\in\bR^{n-1}}\E[dN_{\tau_1}dN_{\tau_2}\dotsm dN_{\tau_{n}}] =K_nd\tau_n$ with $K_n<\infty$.  
We then have 
\eq{
	&\E[|\sprod{j\in\cP_n}f_j(0)|]\\
\leq	&\mat{\int_{[\tau]_1^n\in(-\infty,0)^{n}} \sprod{j\in\cP_n}}|g_j(-\tau_{j-1})|\E[dN_{\tau_1}\dotsm dN_{\tau_{n}}]\\
\leq	&\sprod{j\in\cP_{n-1}}\Lin{g_j}\mat{\int_{[\tau]_1^n\in(0,\infty)^n} }|g_n(\tau_n)|\E[dN_{\tau_1}\dotsm dN_{\tau_{n}}]\\
=	&\sprod{j\in\cP_{n-1}}\Lin{g_j}\mat{ \intoi }|g_n(\tau_n)| K_nd\tau_n\\
=	&K_n\Lwn{g_n}\sprod{j\in\cP_{n-1}}\Lin{g_j}<\infty.
}
Expanding $\E[\lamo(0)^n]$ in terms of $c_0^k$ and $\E[\chio(0)^{n-k}]$ results in 
a weighted sum of $\E[\chio(0)^{n-k}]$ and is, therefore, finite in view of Lemma \ref{lem:ergsta}(a). 

{\it First and second moments: } 
From Lemma \ref{lem:1mom}, 
it follows that $\E[f_j(0)] = \intiom g_j(-u)\E[dN_u] = \Lam\intoi g_j(u)du$, and from Lemma \ref{lem:2mom}(a), we have
\eq{
	\E[f_1(0)f_2(0)] 
=	&\mat{\intiom\intiom} g_1(-v)g_2(-u)\E[dN_vdN_u] \\
= 	&\mat{\intoi\intoi }g_1(v)g_2(u)[C(u-v)+\Lam^2]dvdu.
}

{\it Ergodicity: }Since the processes $\prod_{j\in\cP_n}f_j(t)$ are stationary 
with finite mean, and the counting process $N$ is ergodic by Lemma \ref{lem:erg}, 
by virtue of Birkhoff's ergodic theorem \cite{Pete89}\cite[Chapter12]{Dale08}, 
$\limTi\woT\intoT\prod_{j\in\cP_n} f_j(t)dt\to\E[\prod_{j\in\cP_n}f_j(0)]$ w.p.1. 
\hfill$\square$

{\it Proof of Lemma \ref{lem:dri}. }
For the \ul{first convergence}, we will use Lemma \ref{lem:hmom} to show that 
$\E[|\intoi h_1\star\Delta f_1(t)\prod_{j=2}^n f_j^{A_j}(t)dt|]<\infty$, 
so that the scaling of $\wo{\sqT}$ will ensure the required vanishing w.p.1. 
Define $\rho_j(t,\tau_j) = \intot|h_j(t-v)g_j(v-\tau_j)|dv$ 
with $D_j\trieq\sup_{t,\tau_j}\rho_j(t,\tau_j)\leq\Lin{g_j}\sup_t\intot|h_j(t-v)|dv=\Lin{g_j}\Lwn{h_j}$. 
Putting the absolute signs inside the integrals and changing the order of the integrals, 
we find
\eq{
	&\E[|h_1\star\Delta f_1(t)\sprod{j=2}^n h_j\star f_j^{A_j}(t)|]\\
\leq	&\mat{\int_{A_n}\dotsm\int_{A_2}\intiom }\rho_1(t,\tau_1)\sprod{j=2}^n\rho_j(t,\tau_j)\E[dN_{\tau_1}\dotsm dN_{\tau_n}]\\
\leq	&\sprod{j=2}^nD_j\mat{\int_\bR\dotsm\int_\bR\intiom}\rho_1(t,\tau_1)\E[dN_{\tau_1}\dotsm dN_{\tau_n}]\\
\leq 	&K_n\sprod{j=2}^nD_j\mat{\intio}\rho_1(t,\tau_1)d\tau_1,
}
where 
the last inequality and the constant $K_n$ are from Lemma \ref{lem:hmom}. 
Set $D=K_n\prod_{j=2}^nD_j$ and $S^g_1(t) = \intti |g_1(u)|du$ 
with $\Lwn{S^g_1}=\intoi \intti|g_1(u)|dudt = \intoi\intou dt|g_1(u)|du = \intoi u|g_1(u)|du<\infty$. 
We then have 
$
	\E[|\mat{\intoi} h_1\star\Delta f_1(t)\sprod{j=2}^n f_j^{A_j}(t)dt|]
\leq 	D\mat{\intoi \intot |h_1(t-v)|}$ $\mat{\int_v^\infty} |g_1(\tau_1)|d\tau_1dvdt 
= 	D\Lwn{|h_1|\star S^g_1}=D\Lwn{h}\Lwn{S^g_1}$. 

Now for the \ul{second convergence}, we also show 
$\E[|\intoi h_1\star \Delta f_1(t)dN_t|]<\infty$. 
Change the order of integrals to find 
$\E[|\intoi h_1\star \Delta f_1(t)dN_t|]
\leq\intoi \intiom \rho_1(t,\tau) \E[dN_\tau dN_t]$. 
Since the regions of integrals ensure $\tau\neq t$, 
we have $\E[dN_\tau dN_t] = (\Creg(t-\tau)+\Lam^2)d\tau dt\leq(\Lin{\Creg}+\Lam^2)d\tau dt$ by Lemma \ref{lem:2mom}. 
The finiteness of the expectation follows from $\intoi\intio\rho_1(t,\tau)d\tau dt<\infty$ 
as shown above. 

For the \ul{third convergence}, use $dM_t=dN_t-\lamo(t)dt = dN_t - (c_0+\chio(t))dt$ 
to find $\wo{T^\ep}\intoT h_1\star\Delta f_1(t)dM_t = \wo{T^\ep}\intoT h_1\star \Delta f_1(t) dN_t  - \frac{c_0}{T^\ep}\intoT h_1\star \Delta f_1(t)dt - \wo{T^\ep}\intoT h_1\star \Delta f_1(t)\chio(t)dt\to0$ w.p.1 by the established convergences. 
\hfill$\square$

{\it Proof of Lemma \ref{lem:marint}. }
(a) $f_1(u)$ and $\tilde f_1(u)$ are clearly $\cHtm$-predictable. 
In view of Lemma \ref{lem:mar}(b),  
we only need to check that $\E[\intoT\intitm |g_1(t-v)|dN_v\lamo(t)dt]<\infty$, 
since $\E[\intoT|\tilde f_1(t)|\lamo(t)dt]\leq\E[\intoT|f_1(t)|\lamo(t)dt]\leq\E[\intoT\intitm |g_1(t-v)|dN_v\lamo(t)dt]$.  
However,
\eq{
	&\E[\mat{\intoT|g_1|\star dN_t\lamo(t)dt]}\\
=	&\E[\mat{\intoT|g_1|\star dN_t({c_0+\phio\star dN_t)}dt}]\\
=	&Tc_0\Lam\Lwn{g_1} \\
	&+ T\mat{\intoi\intoi} |g_1(v)|\phio(u)[C(u-v)+\Lam^2]dvdu, 
}
where, by Lemma \ref{lem:ergsta}, the last line follows from stationarity of $|g_1|\star dN_t$ and $(|g_1|\star dN_t)(\phio\star dN_t)$, and is bounded. 

(b) If we show $\woT M_{j,T}^f\to0$, 
then it also follows that $\woT\tilde M_{j,T}^f = \woT M_{j,T}^f-\woT\Delta M_{j,T}^f\to0$ 
since $\woT\Delta M_{j,T}^f\to0$ by Lemma \ref{lem:dri}. 
It is straightforward to verify that both $M^f_j, \tilde M^f_j$ are martingales 
in view of Lemma \ref{lem:mar}(b). 
Then, $\ang{M^f_j}_T =\intoT f_j(t)^2\lamo(t)dt$. 
Since $f_j(t)^2\lamo(t)$ is stationary by Lemma \ref{lem:ergsta} and $\E[f_j(0)^2\lamo(0)]\leq\E[f_j(0)^3]^{2/3}\E[\lamo(0)^3]^{1/3}<\infty$ 
by H\"older's inequality and Lemma \ref{lem:ergsta}(b), 
we can then use Lemma \ref{lem:ergsta}(c) to find $\woT\ang{M^f_j}_T\to \E[f_j(0)^2\lamo(0)]>0$ w.p.1. 
Then $\woT M^f_{j,T} = \frac{M^f_{j,T}}{\ang{M^f_j}_T}\woT{\ang{M^f_j}_T}\to0$, w.p.1 
by the strong law of large numbers (SLLN) for martingales \cite[Corollary 2.6.1]{Lipt89}. 
\hfill$\square$

{\it Proof of Lemma \ref{lem:chiconv}. }
Rewrite $\chihT = \woT\intoT\chitil(t)dt = \woT\intoT\chi(t)dt - \woT\intoT\intiom q(t-u)dN_udt$. 
Under \ref{A2}, 
the first term $\woT\intoT\chi(t)dt\to \E[\chi(0)]= \Lam$ by Lemma \ref{lem:ergsta} 
and Lemma \ref{lem:1mom}. 
Under \ref{A3}, the second term vanishes w.p.1 by Lemma \ref{lem:dri}.
\hfill$\square$

{\it Proof of Lemma \ref{lem:Rconv}.} 
Note that under \ref{A2}, $\chihT\to\E[\chi(0)]=\mu$ w.p.1 by Lemma \ref{lem:chiconv} 
and $\woT\intoT\chi(t)\chi(t)^\top dt\to\E[\chi(0)\chi(0)] = \intoi\intoi q(v)C(u-v)q(u)^\top dvdu + \mu\mu^\top$ w.p.1 by Lemma \ref{lem:ergsta}. 
Subtracting $\woT\intoT\chi(t)\chi(t)^\top dt - \chihT\chihTT$ from $R_T$ 
results in a matrix containing the drift terms possessing the properties as in Lemma \ref{lem:dri} under \ref{A3} and thus vanishing. 
The result for $\RTwo$ also follows in the same way. 
\hfill$\square$

{\it Proof of Lemma \ref{lem:sconv}. }
Both $\VhT=\intoT\chitil(t)dM_t$ and $M_T$ are $\cH_{T-}$-martingales from 
Lemma \ref{lem:marint}(a) and Lemma \ref{lem:mar}(a), respectively. 
Since $\RTwo\to\Rswo$ and $\chihT\to\mu$ w.p.1 by Lemma \ref{lem:Rconv} 
and Lemma \ref{lem:chiconv}, and $\woT\intoT\chitil(t)dM_t\to0$ w.p.1 
by Lemma \ref{lem:marint}(b), 
Lemma \ref{lem:sconv} follows by showing $\frac{M_T}T\to0$ w.p.1, 
which follows the same SLLN for martingales \cite[Corollary 2.6.1]{Lipt89} 
as in the proof of Lemma \ref{lem:marint}(b). 
\hfill$\square$

\myapp{Proofs for Section \ref{sec:clt}}\label{apxC}
This Appendix proves Main Result IV: CLT under Misspecification. 
We derive martingale representations for the bias terms $\BaT$ and $\BcT$ and apply the functional martingale CLT (Lemma \ref{lem:mclt}). 
The primary task is to isolate the martingale core by proving that the bias and remainder terms vanish as $T \to \infty$. 
These vanishing proofs are lengthy. 
To better present the proofs, we provide some general results 
and define intermediate variables and operators 
that do not persist in the final result.

We define the tail integral operator $\operatorname{S}$. For a measurable function 
$g$, let $(\Sr g)(t) \trieq \intti |g(u)|du, t\geq0$. 
For simplicity, we denote the resulting function 
$S^g(t)\trieq (\Sr g)(t)$. For indexed scalar functions $g_j$, we adopt the simplified notation $S^g_j(t)\trieq (\Sr g_j)(t)$. %
We will also write $(\Sr g_j^k)(t) \trieq \intti |g_j(u)|^kdu$. 
If further $\intoi t|g_j(t)|dt<\infty$, 
the standard moment identity (see proof of Lemma \ref{lem:dri})
\eqn{
\Lwn{S^g_j}
=\mat{\intoi \intti|g_j(u)|dudt = \intoi u|g_j(u)|du}, \label{eq:Snorm}
}
ensures that $\Lwn{S^g_j}<\infty$. 
We further require the following general results.

\Blem{lem:Sint}
Let $g_1,g_2\in\Lw\cap\Li$ with $\intoi u|g_j(u)|du<\infty$. 
Then, for $k\in\bN$, 
\nlist{
\ita $\Lwn{\Sr g_1^k}<\infty, \ \Lwn{\Sr g_1\star g_2}<\infty, \ \Lwn{\Sr g_1^{\star k}}<\infty.$
\itb Specially, under \ref{A3}, \\
$
\Lwn{\Sr \phio^k}<\infty, \quad \Lwn{\Sr q_j^k}<\infty, \quad \Lwn{\Sr \kap^k}<\infty.
$
}
\Elem
\pro{(a) By the moment identify \eqref{eq:Snorm}, 
$\Lwn{\Sr g_j^k}=\intoi u|g_j(u)|^kdu\leq\Lin{g_j}^{k-1}\intoi u|g_j(u)|du<\infty$. 
We also have 
$
	\Lwn{\Sr g_j\star g_{j'}} 
= 	\mat{\intoi }u|g_j\star g_{j'}(u)|du
\leq 	\mat{\intoi \intou }u|g_j(u-v)| |g_{j'}(v)|dvdu
=	\mat{\intoi \intou }(u-v)|g_j(u-v)| |g_{j'}(v)|dvdu
	 + \mat{\intoi \intou }|g_j(u-v)| v|g_{j'}(v)|dvdu.
$
Observing the convolutional structure and using Young's inequality, 
we find $\Lwn{\Sr g_j\star g_{j'}}\leq \Lwn{g_{j'}}\intoi u|g_j(u)|du+\Lwn{g_j}\intoi u|g_{j'}(u)|du=\Lwn{g_j}\Lwn{S^g_{j'}}<\infty$. 

Repeat the above recursively and notice the identity  $\Lwn{g_j^{\star k}}=\Lwn{g_j}^k$ 
to find 
\eq{
	&\Lwn{\Sr g_j^{\star k}} 
= 	\Lwn{\Sr g_j\star g_j^{\star(k-1)}} \\
\leq 	&\Lwn{g_{j}}\mat{\Lwn{\Sr g_j^{\star(k-1)}}+\Lwn{g_j}^{k-1}\intoi u|g_{j}(u)|du}\\
\leq	&\dotsm\leq k\Lwn{g_j}^{k-1}\mat{\intoi }u|g_j(u)|du=k\Lwn{g_j}^{k-1}\Lwn{S^g_j}.
}

(b) The boundedness of $\Lwn{\Sr \phio^k}$ and $\Lwn{\Sr q_j^k}$ 
is clear from part(a). 
We prove for the Hawkes resolvents. Since $\kap\in\Lw\cap\Li$, 
it suffices to show $\Lwn{\Sr \kap}<\infty$. 
Note that $\Lwn{\Sr \kap}=\intoi u\snwi\phio^{\star n}(u)du$. 
Recursively use Minkowski's norm inequality \cite{Foll99}
to find 
$
	\Lwn{\Sr \kap}
\leq	\mat{\intoi u\phio(u)dt + \intoi u\ssum{n=2}^\infty\phio^{\star n}(u)du}
\leq	\dotsm \leq\ssum{n=1}^\infty\mat{\intoi} u\phio^{\star n}(u)du = \ssum{n=1}^\infty\Lwn{\Sr \phio^{\star n}}$. 
From the previous proof, we have $\Lwn{\Sr \phio^{\star n}}\leq n\Lwn{\phio}^{n-1}\Lwn{\Sphio} = n\Gam^{n-1}\Lwn{\Sphio}$. 
Thus, $\Lwn{\Sr \kap}\leq \ssum{n=1}^\infty n\Gam^{n-1}\Lwn{\Sphio}
=\wo{(1-\Gam)^2}\Lwn{\Sphio}<\infty$.
}

\Blem{lem:Svan}
Let $N$ be stationary satisfying \ref{A1} and $g_1,g_2\in\Lw\cap\Li$ be deterministic functions satisfying $\intoi t |g_j(t)|dt<\infty$. For any $\ep>0$, as $\Ttoi$, 
\eq{
&\mat{\wo{T^\ep} S^g_j\star d\NtilT, \ \wo{T^\ep} S^g_j\star dM_T, \ \wo{T^\ep}\intoT S^g_1(t) \intot g_2(u)dudt} \\
&\mat{\wo{T^\ep} \intoT S^g_1(t) (g_2\star d\Ntilt )dt, \quad \wo{T^\ep} \intoT S^g_1(t) (g_2\star dM_t )dt,}
}
all converge to $0$ w.p.1.
\Elem
\pro{
Simply take the absolute expectations of the integrals to find that they are all bounded, e.g. $\E[|\intoi S^g_1(t)(g_2\star dM_t)dt|]\leq\intoi S^g_1(t) \E[|g_2|\star d\Ntilt] dt+\intoi S^g_1(t) \E[|g_2|\star \lamo(t)]dt = 2\Lam\intoi S^g_1(t)\intot |g_2(u)|dudt\leq 2\Lam\Lwn{g_2}\Lwn{S^g_1}<\infty$. Then scaling by $\wo{T^\ep}$ establishes the required convergence. 
}

{\it Proof of Lemma \ref{lem:bct}. }
Change the order of integrals to find
\eqn{
	\chihT-\mu	
=	&\mat{\woT\intoT\intotm} q(t-u)dN_u dt - \Lam \bbw{P}\nonumber\\
=	&\mat{\woT}\mat{\intoT\int_u^T} q(t-u)dt dN_u-\Lam\bbw{P}\nonumber\\
= 	&\mat{\woT}\mat{\intoT} (\bbw{P}-\Sq(T-u))dN_u-\Lam\bbw{P}\nonumber\\
=	&(\LamhT-\Lam)\bbw{P}+o_p(\bbw{P}),\label{eq:chidif}
}
where the $o_p(\bbw{P})$ term is $ - \woT \Sq\star d\NtilT$ thanks to Lemma \ref{lem:Svan}. 
We thus have $\BcT = \sqrt{T}(\LamhT-\Lam)-\sqrt{T}(\chihT-\mu)^\top\alps = 
\sqrt{T}(\LamhT-\Lam)(1-\Gams)+o_p(\bbw P)$. 
But $\sqrt{T}(\LamhT-\Lam) = \wo{\sqrt{T}}\intoT (dN_t-\Lam dt)=\wo{\sqrt{T}}(M_T+\intoT(\lamo(t)-\Lam)dt = \wo{\sqrt{T}}(M_T + \intoT\kap\star dM_tdt+\intoT\zeta\star\eta(t)dt)$, by Lemma \ref{lem:lamm}, 
and further $\wo{\sqrt{T}}\intoT\kap\star dM_t dt = \wo{\sqrt{T}} \intoT\int_0^{T-u}\kap(t)dtdM_u = 
\wo{\sqrt{T}}\intoT\intoi\kap(t)dtdM_u - \wo{\sqrt{T}}\intoT\int_{T-u}^\infty \kap(t)dtdM_u = \frac{\Gam}{1-\Gam}\frac{M_T}{\sqrt{T}}-\wo{\sqrt{T}} \Skap\star M_T$. 
By Lemma \ref{lem:Sint}(b) and Lemma \ref{lem:Svan}, we have $\wo{\sqrt{T}} \Skap\star dM_T\convp0$. 
Thus, $\BcT = \wo{\sqrt{T}}(\frac{\Gam}{1-\Gam}+1)(1-\Gams)M_T+o_p(1)=\wo{\sqrt{T}}\frac{1-\Gams}{1-\Gam}M_T+o_p(1)$.
\hfill$\square$

{\it Proof of Lemma \ref{lem:bat}. }
We establish the result through a four-stage decomposition. 
For clarity, we introduce the intermediate terms $U_T$, $Y_T$, and $Z_T$:
\eq{
	&\mat{U_T
=	& \wo{\sqrt{T}}\intoT\intot q(t-v)\nu(dv)\intot\Dphi(t-u)\nu(du)dt}\\
	&\mat{Y_T
=	& \wo{\sqrt{T}}\intoT\intoum W_T(u,v)+W_T(v,u)dM_vdM_u}\\
	&\mat{Z_T
=	& \wo{\sqrt{T}}\intoT\intoum W(u-v) dM_vdM_u,}
}
where the $T$-dependent kernel $W_T(u,v)$ is defined as 
\eq{
	W_T(u,v) 
=	&\mat{\int_{\max\{u,v\}}^T}a(t-v)b(t-u)dt\\
=	& \mat{\intoT} a(t-v)b(t-u)dt,
}
where $a(t) = q\star\zeta(t)$ and $b(t) = \Dphi\star\zeta(t)$, 
and $T$-invariant two-sided kernel $W$ as defined in the theorem 
is $W(u-v) =\mat{ \limTi W_T(u,v)+W_T(v,u)=W(v-u)}$.

We denote $a_j(t)=q_j\star\zeta(t), W_{j,T}(u,v)=\intoT a_j(t-v)b(t-u)dt$ 
and $W_j(u-v)$ 
to be the $j$-the element of $a(t)$, $W_T(u,v)$ and $W(u-v)$, respectively. 
We have 
\eqn{
a_j,b\in\Lw\cap\Li, \label{eq:abL}
}because by Young's inequality for any $1\leq p\leq\infty$, 
$\Lpn{a_j}\leq\Lpn{q_j}\Lwn{\zeta}<\infty$ 
and $\Lpn{b}\leq\Lpn{\Dphi}\Lwn{\zeta}<\infty$. 
Also, 
\eqn{
&\Lwn{S^a_j}<\infty, \quad, \Lwn{S^b}<\infty, \label{eq:Sab}
}
because $\Lwn{S^a_j}=\intoi u|q_j\star\zeta(u)|du = \intoi u|q_j(u)|du + \intoi u|q_j\star\kap(u)|du<\infty$ 
and $\Lwn{S^b}=\intoi u|\Dphi\star\zeta(u)|du = \intoi u|\Dphi(u)|du + \intoi u|\Dphi\star\kap(u)|du<\infty$, in view of Lemma \ref{lem:Sint} and \eqref{eq:tphi}.

The proof proceeds as follows.  
[Stage 1] Centered measure approximation: We first express the bias $\BaT$ in terms of the centered measure $\nu$, showing that $\BaT = U_T + o_p(1)$. 
Replacing the centered measure by the martingale measure developed in 
Lemma \ref{lem:lamm} will ensure that the bias terms all vanish. 
[Stage 2] Predictability: To allow for martingale calculus, we approximate $U_T$ by the double martingale integral $Y_T$, establishing that $U_T = Y_T + o_p(1)$ where the integrand is $\cH_{u-}$-predictable.
[Stage 3] Martingale representation: Finally, we remove the $T$-dependence of the kernel by showing $Y_T=Z_T+o_p(1)$. We prove that $Z_T$ is a valid martingale (after scaling), yielding the required representation $\BaT = Z_T + o_p(1)$. 
[Stage 4] Centered measure recovery: By applying a reverse \mr\ of the intensity deviation, we recover $W\star dM_u$ 
in the form of $\nu(du)$ to match the quoted result, on which we can apply the 
functional martingale CLT.

{\bf [Stage 1]} Rewrite 
\eq{
	\BaT 
= 	&\mat{\sqrt{T}(\RTwo-\RT\Rs^{-1}\Rswo) 
= 	\sqrt{T}(\RTwo-\RT\alps)}\\
= 	&\mat{\wo{\sqrt{T}}\intoT(\chitil(t)-\chihT)(\chio(t)-\chitil(t)^\top\alps)dt} \\
= 	&\mat{\wo{\sqrt{T}}\intoT(\chitil(t)-\chihT)(\chitilo-\chitil(t)^\top\alps)dt+o_p(\bbw{P})}\\
=	&\mat{\wo{\sqrt{T}}\intoT(\chitil(t)-\chihT)(\Dphi\star d\Ntilt)dt+o_p(\bbw{P}),}
} 
where the $o_p(\bbw{P})$ term is $\wo{\sqrt{T}}\intoT(\chitil(t)-\chihT)\Dchio(t)^\top dt\alps$ thanks to Lemma \ref{lem:dri}. 

Subtracting $\BaT$ from $U_T$ and using the identifies
\eq{
&\mat{\mu=\Lam\intit q(t-v)dv=\Lam \intot q(t-v)dv+\Lam\Sq(t)}\\
&\mat{\Gam = \intit \phio(t-v)dv = \intot \phio(t-v)dv + \Sphio(t)}\\
&\mat{\Gams = \intit \phis(t-v)dv = \intot\phis(t-v)dv + \intti\phis(v)dv},
}
we find 
\eq{
	&U_T - \BaT\\
=	&\mat{\sqT(\chihT-\mu)\bra{\woT\intoT\Dphi\star d\Ntilt dt - \Lam(\Gam-\Gams)}}\\
	&\mat{+ \frac\Lam{\sqT}\intoT(\chitil(t)-\mu)\bra{\Sphio(t)-\intti\phis(v)dv}dt} \\
	&\mat{+ \frac\Lam{\sqT}\intoT\Sq(t)\intotm\Dphi(t-u)\nu(du)dt + o_p(\bbw P)}.\\
}
The last two $\Sr$-related terms\footnote{Note that $\phis(t)$ is not guaranteed to be nonnegative for $t\geq0$, so we cannot equate $\intti\phis(v)dv$ with $\Sphis(t)=\intti|\phis(v)|dv$. Nevertheless, because absolute bounds are sufficient to show the term vanishes, we can still apply the $S$-properties.} vanish w.p.1, because of Lemma \ref{lem:Svan}. 
For the first term, since $\woT\intoT \Dphi\star d\Ntilt dt\to \Lam(\Gam-\Gams)$ 
by Lemma \ref{lem:ergsta} and Lemma \ref{lem:dri} 
and $\sqrt{T}(\chihT-\mu)=(\LamhT-\Lam)\bbw P+o_p(\bbw P) \RA\cN(0,\frac\Lam{(1-\Gam)^2}\bbw P\bbw P^\top)$ from \eqref{eq:chidif} and 
a standard CLT result \cite{Bacr13}, 
use Slutsky's theorem \cite{Van00} to find that the first term also vanish w.p.1. 
We thus find $\BaT=U_T+o_p(\bbw P)$.

\noindent{\bf [Stage 2]} We can now apply Lemma \ref{lem:lamm} 
to replace the centered measure $\nu$ with the martingale measure $M$ in $U_T$. 
Using Lemma \ref{lem:lamm}(b) 
and changing the order of integrals, 
we find 
\eqn{
	U_T
=	&\mat{\wo{\sqT}\intoT\intotm q(t-v)\nu(dv)\intotm\Dphi(t-u)\nu(du)dt}\nonumber\\
=	&\mat{\wo{\sqT}\intoT (a\star dM_t)(b\star dM_t)dt}\label{eq:U1}\\
	&\mat{+ \wo{\sqT}\intoT a\star\eta(t)b\star\eta(t)dt}\label{eq:V1}\\
	&\mat{+\wo{\sqT}\intoT a\star\eta(t) (b\star dM_t)dt}\label{eq:V2}\\
	&\mat{+\wo{\sqT}\intoT (a\star dM_t) b\star\eta(t)dt}.\label{eq:V3}
}
We will show that \eqref{eq:V1}-\eqref{eq:V3} all vanish asymptotically and 
\eqref{eq:U1} is equal to $X_T+o_p(\bbw P)$.

$\ang{\mbox{Vanishing of \eqref{eq:V1}-\eqref{eq:V3}}}$  
Lemma \ref{lem:dri} implies that terms containing
$\eta(t)$ vanish. 
We adapt the argument to account for the centered measure 
and the martingale increaments. 

For \eqref{eq:V1}, note that $\eta(t) =\Dphio(t)-\Lam\Sphio(t)$. 
\eqref{eq:V1} splits into four terms: 
$\wo{\sqT}\intoT a\star \Dphio(t)b\star\Dphio(t)dt 
-\frac{\Lam}{\sqT}\intoT a\star\Dphio(t)b\star\Sphio(t)dt
-\frac{\Lam}{\sqT}\intoT a\star\Sphio(t)b\star\Dphio(t)dt
+\frac{\Lam^2}{\sqT}\intoT a\star\Sphio(t)b\star\Sphio(t)dt$.
The first term converges to $0$ w.p.1 as covered in Lemma \ref{lem:dri}. 
The last term is deterministic and tends to $0$ 
because by \CSi\ followed by Young's inequality, we have $\intoi a_j\star\Sphio(t)b\star\Sphio(t)dt 
\leq\Ltn{a_j\star\Sphio}\Ltn{b\star\Sphio}\leq \Ltn{a_j}\Ltn{b}\Lwn{\Sphio}^2<\infty$. 
The remaining cross-terms vanish by bounding the deterministic factors and applying Lemma \ref{lem:dri} to the remaining stochastic integrals.

For \eqref{eq:V2} and \eqref{eq:V3}, use $dM_t = dN_t - \lamo(t)dt$ to 
expand, e.g., $b\star dM_t =b\star d\Ntilt - c_0 \intot b(u)du - \intot b(t-u)\chio(u)du$ 
and apply Lemma \ref{lem:dri} to establish the required convergence.

$\ang{\mbox{Equivalence of \eqref{eq:U1}}}$ 
Exchange the order of integrals in \eqref{eq:U1} to find 
$\intoT a\star dM_tb\star dM_tdt=\intoT(\intotm a(t-v)dM_v)(\intotm b(t-u)dM_u)dt 
= \intoT\intoT\int_{\max\{u,v\}}^T a(t-v)b(t-u)dtdM_vdM_u
= \intoT\intoT\intoT a(t-v)b(t-u)dtdM_vdM_u
=\intoT\intoT W_T(u,v)dM_vdM_u$. 
Writing \eqref{eq:U1} in a causal form will result in a diagonal term due to 
the jumping nature of the process: 
\eq{
	U_T 
=	&\mat{\wo{\sqT}\intoT W_T(u,u)dN_u + Y_T,} 
}
where $Y_T = \intoT\intotm W_T(u,v)+W_T(v,u)dM_vdM_u$, as defined before. 
We now show the diagonal term $\wo{\sqT}\intoT W_T(u,u)dN_u\to0$ w.p.1 so that $U_T=Y_T+o_p(\bbw P)$ as required. 
We will show $\E[|\intoi W_{j,T}(u,u)dN_u|]$ $\to0$, then 
scaling of $\wo{\sqT}$ ensures the required vanishing. 
Set $e(u)=a(u)b(u)$, 
and denote $e_j(u)=a_j(u)b(u)$ as 
the $j$-th element of $e(u)$. 
Then, 
\eq{
	&\E[\bigl|\mat{\intoi} W_{j,T}(u,u)dN_u\bigr|]
\leq	\Lam\mat{\intoi|\int_0^{T-u}}e_j(t)dt|du\\
=	&\Lam\mat{\intoi}\bigl|\mat{\intoi} e_j(t)dt - \mat{\int_{T-u}^\infty }e_j(t)dt\bigr|du.
}
Below, we will show (a) $\intoi e(u)du=0$ and (b) $e\in \Lw\cap\Li$ with $\intoi u|e(u)|du<\infty$, so that 
$\E[|\intoi W_{j,T}(u,u)dN_u|]\leq \Lam\intoi\int_{T-u}^\infty |e_j(t)|dtdu
= \Lam\intoi S^e_j(T-u)du\leq\Lam \Lwn{S^e_j}<\infty$, by Lemma \ref{lem:Sint}.

(a) Note that $\zeta$ has FT $\bar\zeta(\jw) = \wo{1-\phibo(\jw)}$ 
(as discussed upon its definition in Section \ref{sec:clt}.3) and 
$\bar C(\jw) = \frac{\Lam}{|1-\phibo(\jw)|^2}$ (in Lemma \ref{lem:2mom}(b)). 
Also recall the $\Rs,\Rswo$ expressions in Lemma \ref{lem:Rconv}(b), 
the pseudo-true value $\alps=\Rs^{-1}\Rswo$ in Theorem \ref{thm:theconv}, 
and the pseudo-true HIR $\phis(t)=q(t)^\top\alps$. 
By Parseval's theorem,
\eq{
	&\Lam\mat{\intoi }e(u)dt\\
=	&\mat{\frac\Lam{2\pi}\intii } \bar q(\jw)\bar\zeta(\jw)(\phibo(-\jw)-\phibs(-\jw))\bar\zeta(-\jw)d\omega\\
=	&\mat{\wo{2\pi}}\mat{\intii  \bar q(\jw)(\phibo(-\jw)-\bar q(-\jw)^\top\alps)\frac{\Lam}{|1-\phibo(\jw)|^2}d\omega}\\
=	&\mat{\wo{2\pi}}\mat{\intii }\bar q(\jw)(\phibo(-\jw)-\bar q(-\jw)^\top\alps)\bar C(\omega)d\omega\\
=	&\Rswo-\Rs\alps=0.
}

(b) For any $1\leq p\leq\infty$, use successively the \CSi\ and Young's inequality to find
$\Lpn{e}\leq \Ver{q_j\star\zeta}_{L^{2p}}\Ver{\Dphi\star\zeta}_{L^{2p}}
\leq \Lwn{\zeta}^2\Ver{q_j}_{L^{2p-1}}\Ver{\Dphi}_{L^{2p-1}}<\infty$. 
Further, $\intoi u|e_j(u)|du =\intoi uq_j\star\zeta(u)|\Dphi\star\zeta(u)|du\leq\Lin{q_j\star\zeta}\intoi u(|\Dphi(u)|+|\Dphi\star\kap(u)|du$. 
Since $\intoi u|\Dphi(u)|dt<\infty$ (see \eqref{eq:tphi}), $\intoi u\kap(u)du<\infty$ by Lemma \ref{lem:Sint}, and $\Lin{q_j\star\zeta}\leq\Lin{q_j}\Lwn{\zeta}<\infty$ by Young's inequality, condition (b) is satisfied. 
Therefore, both conditions (a) and (b) are satisfied to conclude $U_T=Y_T+o_p(\bbw P)$. 

\noindent{\bf [Stage 3]} 
The term $Y_T$ possesses the requisite double-integral structure with respect to the martingale increments, where a deterministic kernel $W_T$ and 
the inner integration limit $u-$ ensure predictability. However, the kernel's dependence on the time horizon $T$ precludes a direct application of 
the ergodic Lemma \ref{lem:ergsta}. 

Following some straightforward calculations, one can find 
$W(u-v)-W_T(u,v)-W_T(v,u)= \DW(u,v)+\DW(v,u)$, 
where $\DW(u,v)=\intTi q\star\zeta(t-v)\Dphi\star\zeta(t-u)dt$. 
We emphasize that $W$ is defined on $\bR$. 
We will show $Y_T=Z_T+o_p(\bbw P)$ by showing 
first (c) $W$ exists on $\bR$, 
so $Z_T$ is well-defined, and then (d) 
$Z_T-Y_T = \wo{\sqT}\intoT\intoum \DW(u,v)+\DW(v,u)dM_vdM_u\convp0$.

(c) $W$ is well-defined if $\Lin{W}<\infty$. 
We show a stronger result: $\Lpn{W}<\infty$, for all $1\leq p\leq\infty$. 
Set $W_{1,j}(u) = \intoi a_j(t+u)b(t)dt$ 
and $W_{2,j}(u) = \intoi a_j(t)b(t+u)dt$, 
so the $j$-th element of $W(u)$ is $W_j(u) = W_{1,j}(u)+W_{2,j}(u)$. 
We also set $\check a(t) = a(-t), \check a_j(t) = a_j(-t)$ and $\check b(t)=b(-t)$. 
We first write $W(u)$ in a convolutional form:
\eq{
	&W_{1,j} (u)
=	\mat{\intoi } a_j(t+u)b(t)dt
=	\mat{\intoi }a_j(t+u)\check b(-t)dt\\
=	&\mat{\intio} a_j(u-t)\check b(t)dt
=	\mat{\int_{\bR}} a_j(u-t)\check b(t)dt
=	a_j\star \check b(u),
}
where the second last equality follows from $\check b(t)=0,t>0$. 
Similarly $W_{2,j}(u)= \check a_j\star b(u)$. 
Then, by Young's inequality
$
\Lpn{W_{1,j}}\leq\Lwn{\check b}\Lpn{a_j}=\Lwn{b}\Lpn{a_j}<\infty$, and
$\Lpn{W_{2,j}}\leq\Lwn{b}\Lpn{\check a_j}=\Lwn{b}\Lpn{a_j}<\infty.
$
We thus have $\Lpn{W_j}\leq\Lpn{W_{1,j}}+\Lpn{W_{2,j}}<\infty$, 
for any $1\leq p\leq \infty$, by Minkowski's norm inequality \cite{Foll99}. 

(d) We will show $\wo{\sqT}\intoT\intoum\Delta W_T(u,v)dM_vdM_u\convp0$ using 
Lemma \ref{lem:mclt}. In this instance, it suffices to verify conditions Lemma \ref{lem:mclt}(a,b). Since Lemma \ref{lem:mclt}(b) establishes that the predictable quadratic variation converges to zero (i.e., $\Sigma = 0$), the Lindeberg condition is satisfied as a direct consequence. 
Define 
$
\Vs = \mat{\wo{\sqT}\intosum\DW(u,v)dM_v, s\in[0,1]}$ and $
X_\tau^T = \mat{\intotT \VTu_1dM_u, \tau\in[0,1]} $.
Also set $\DWj(u,v)=\intTi a_j(t-v)b(t-u)dt$ 
as the $j$-th element of $\DW(u,v)$ and $\VTu_{j,s}=\wo{\sqT}\intoum\DWj(u,v)dM_v$ 
as the $j$-th element of $\Vs$. 

$V^{T,u}_{j,s}$ is a square-integratable $\cH_{su-}$-martingale (by Lemma \ref{lem:mclt}(a)) because 
\eqn{
	&\E[\mat{\intosu} \DWj(u,v)^2\lamo(v)dv]=\Lam\mat{\intosu}\DWj(u,v)^2dv\nonumber\\
= 	&\Lam\mat{\intosu}\bra{\mat{\intTi }a_j(t-v)b(t-u)dt}^2dv\nonumber\\
\leq	&\Lam\mat{\intosu\intTi} a_j(t-v)^2dt \mat{\intTi} b(t-u)^2dtdv\label{eq:DW2}\\
\leq	&\Lam\Ltn{b}^2\Ltn{a_j}^2su<\infty,\label{eq:DW}
}
where the third line uses \CSi. 

We now show $X_\tau^T$ is a square-integrable $\cH_{\tau T-}$-martingale. 
First, by H\"older's inequality and then the \BDGi \cite{Prot04}, 
\eq{
	&\E[\mat{\intotT}(\VTu_{j,1})^2\lamo(u)du]\\
\leq	&\mat{\intotT}\E[(\VTu_{j,1})^3]^{\frac23}\E[\lamo(u)^3]^{\frac13}du\\
=	&\E[\lamo(0)^3]^{\frac13}\mat{\intotT}\E[(\VTu_{j,1})^3]^{\frac23}du\\
\leq	&D\E[\lamo(0)^3]^{\frac13}\mat{\intotT}\E[\ang{\VTu_j}_1^{\frac32}]^{\frac23}du,
}
where we used H\"older's inequality in the first line, 
stationarity (Lemma \ref{lem:ergsta}(a)) in the second line, 
and the \BDGi \cite{Prot04} in the last line with some constant $D$. 
However, by Minkowski's integral inequality \cite{Foll99},
\eq{
	\E[\ang{\VTu_j}_1^{\frac32}]^{\frac23}
=	&\mat{\woT\E[(\mat{\intou} \wo{v}v\DWj(u,v)^2\lamo(v)dv)^{\frac32}]^{\frac23}}\\
\leq	&\mat{\woT\intou \wo v(\E[(v\DWj(u,v)^2\lamo(v))^{\frac32}])^{\frac23}dv}\\
=	&\mat{\frac{\E[\lamo(0)^{\frac32}]^{\frac23}}{T}\intou \DWj(u,v)^2dv.}
}
From \eqref{eq:DW}, $\E[\ang{V_j^T}_1^{\frac32}]^{\frac23}\leq T^{-1}{u\Lam \Ltn{b}^2\Ltn{a_j}^2}<\infty\Ra\E[\intotT(\VTu_{j,1})^2\lamo(u)du]<\infty$. 

$\ang{\mbox{Check Lemma \ref{lem:mclt}(b)}}$ 
By Markov's inequality, the quadratic variation $\ang{X^T}_\tau = \woT\intotT(\VTu_1)(\VTu_1)^\top\lamo(u)du$ 
will converge in probability to $0$ if its expectation vanishes. 
It suffices to show that $\woT\E[\intotT(\VTu_{j,1})^2\lamo(u)du]\to0$. 
By the above inequalities, we only require \\ 
$\intoi\intou\DWj(u,v)^2dvdu<\infty$. 
However, from \eqref{eq:DW2}, 
\eq{
	&\mat{\intoT\intou}\DWj(u,v)^2dvdu\\
\leq	&\mat{\intoT\intou}\bra{\mat{\intTi }a_j(t-v)^2dt}\bra{\mat{\intTi }b(t-u)^2dt}dvdu\\
=	&\mat{\intoT\intou}\bra{\mat{\int_{T-v}^\infty} a_j(t)^2dt}\bra{\mat{\int_{T-u}^\infty }b(t)^2dt}dvdu\\
\leq	&\mat{\intoT\intoT}\bra{\mat{\int_{T-v}^\infty} a_j(t)^2dt}\bra{\mat{\int_{T-u}^\infty }b(t)^2dt}dvdu\\
=	&\mat{\intoT\int_{v}^\infty} a_j(t)^2dtdv\mat{\intoT\int_{u}^\infty} b(t)^2dtdu\\
\leq	&\Lwn{\Sr a^2_j}\Lwn{\Sr b^2}<\infty
}
where the first inequality follows from \CSi\ and the finiteness 
follows from \eqref{eq:Sab}. 
Therefore, $\wo{\sqT}\intoT\intoum\DW(u,v)dM_vdM_u\convp0$. 
We omit the proof of $\wo{\sqT}\intoT\intoum\DW(v,u)dM_vdM_u\convp0$ 
because of its symmetry. 

\noindent{\bf [Stage 4]} We now have a desirable structure 
$\BaT = \wo{\sqT}\intoT \intoum W(u-v)dM_vdM_u+o_p(\bbw P)$. 
We recover the inner integral to a centered measure representation 
so that we can apply the functional CLT. 
Recall the convolutional structure $W(u) = a\star\check b(u)+\check a\star b(u)$  
from [Stage 3](c). 
Define the truncated causal kernels $\Wtil(u) = W(u)\cI_{u\geq0}$ 
 and 
$h^\alp(u) = \Wtil(u) - \Wtil\star\phio(u)$. 
Then, use $\zeta(u) = \delta(u)+\kap(u)=\delta(u)+\phio\star\zeta(u)$ and Lemma \ref{lem:lamm} to find 
$
	\mat{\intoum} W(u-v)dM_v 
= 	\Wtil\star dM_u 
= 	\Wtil\star\zeta\star dM_u - \Wtil\star\phio\star\zeta\star dM_u
=	\mat{\intoum} h^\alp(u-v)\nu(dv) - h^\alp\star\zeta\star\eta(u)$. 
We thus have
$
	\BaT
=	\mat{\wo{\sqT}\intoT}(\chitilha(t)-\muha)dM_u
	+o_p(\bbw P),
$
as in the Lemma, where the $o_p(\bbw P)$ term is $-\wo{\sqT}\intoT h^\alp\star\zeta\star\eta(u)dM_u+\frac\Lam{\sqT}(\Sr \ha)\star dM_T$. 
The vanishing of the first term follows from $\Lwn{\ha_j\star\zeta}<\infty$ and applying Lemma \ref{lem:dri}. The second term vanishes thanks to Lemma \ref{lem:Svan}.

Note that $\Lpn{W_j}<\infty,1\leq p\leq\infty\Ra\Wtil_j\in\Lw\cap\Li$ from [Stage 3](c) 
and $\Lwn{S^a_j}<\infty, \Lwn{S^b}<\infty\Ra\intoi u|\Wtil_j(u)|du<\infty$ from \eqref{eq:Sab} and Lemma \ref{lem:Sint}. 
We have $\ha_j\in\Lw\cap\Li$ and $\intoi u|\ha_j(u)|du<\infty$, as quoted. 
\hfill$\square$

\bibliographystyle{myplain}
\bibliography{LSH}  

\noindent
{\bf **Supplementary Material}\\
{\it I. Derivations of the CRB inequality in Section 6.3}

Use successively Jensen's inequality and the \CSi, we have 
\eq{
(y^\top \Gs x)^2&\leq\E[\bra{y^\top\wo{\sqrt{\lamo(0)}}\xi(0)\xi(0)^\top\sqrt{\lamo(0)}x}^2]\\
	&\leq(y^\top\Sigcrb^{-1} y)(x^\top\Gs\Sigoth\Gs x),
}
for any $x,y\in\bR^{P+1}$. Then set $y=\Sigcrb u$ and $x=\Gs^{-1}u$ to find 
\eq{
	&(u^\top \Sigcrb u)^2\leq(u^\top\Sigcrb u)(u^\top \Sigoth u)\\
\Ra &u^\top(\Sigcrb-\Sigoth)u\leq0,}
for any $u\neq0\in\bR^{P+1}$.

\noindent
{\bf II. Computations of the Pseudo-True Values in Section 7.1}

{\bf Computation of $\Rs,\Rswo$:} 
Given their spectral integrals (5.2), (5.3), 
the explicit LTs $\bar q_j(s)$ and $\bar p_k(s)$, 
and the chosen frequency range $[-N_{\ome}\dome,N_{\ome}\dome]$ with grid width $\dome$, 
we can estimate $\Rs$ and $\Rswo$ as
\eq{
	&R_* \approx \frac{\dome}{2\pi}\ssum{n=-N_\ome}^{N_\ome}\frac{\bar q(\jmath n\dome)\bar q(-\jmath n\dome)^\top}{|1-\phibo(\jmath n\dome)|^2}\\
	&R_*^{(1,0)} \approx \frac{\dome}{2\pi}\ssum{n=-N_\ome}^{N_\ome}\frac{\bar q(\jmath n\dome)\phibo(-\jmath n\dome)}{|1-\phibo(\jmath n\dome))|^2},}
where $\phibo(s)=\alpo^\top\bar p(s)$. 

{\bf Compution of the \pt\ parameters $\alps,\cs$: } 
Directly, $\alp_* = R_*^{-1}  R_*^{(1,0)}$. Then the \pt\ branching ratio $\Gams = \alps^\top\bbw P$ 
and the \pt\ background rate $c_* = \Lam(1-\Gams)$. 

{\bf Compution of $\muha$: } 
Note that $\muha = \Lam\intoi\ha(u)du = \Lam(1-\Gam)\intoi \Wtil(u)du$. 
We first find $\intoi\Wtil(u)du= \intoi\intoi a(t+u)b(t)dtdu + \intoi \intoi a(t)b(t+u)dtdu 
= \intoi\intti a(u)dub(t)dt + \intoi\intti b(u)dua(t)dt$. 
By Parsavel's theorem, we can estimate $\muha$ as
\eq{
	\mu_h^\alpha 
\approx	&\Lam(1-\Gam)\frac{\dome}{2\pi}\ssum{n=-N_\ome}^{N_\ome} \frac{\frac{\bbw P}{1-\Gam}-\bar a(\jmath n\dome)}{\jmath n\dome}\bar b(-\jmath n\dome)\\
	&+\Lam(1-\Gam)\frac{\dome}{2\pi}\ssum{n=-N_\ome}^{N_\ome}\frac{\frac{\Gam-\Gams}{1-\Gam}-\bar b(\jmath n\dome)}{\jmath n\dome}\bar a(-\jmath n\dome),
}
where $\bar a(s) = \frac{\bar q(s)}{1-\phibo(s)}, \bar b(s) = \frac{\Dphib(s)}{1-\phibo(s)}, \Dphib(s) = \phibo(s)-\alps^\top\bar q(s)$. 
Specially at $n=0$, the direct calculation results in sigularity. 
However, note that, 
\eq{
A_0\trieq&\limso\frac{\frac{\bbw P}{1-\Gam}-\bar a(s)}{s} 
= 	\limso\frac\partial{\partial s}\bra{-{\bar a(s)}} \\
= 	&\limso\frac{\frac{\partial}{\partial s}(-\bar q(s))(1-\phibo(s))+\frac{\partial}{\partial s}(-\phibo(s))\bar q(s)}{(1-\phibo(s))^2}\\
=	&\frac{\limso[\frac{\partial}{\partial s}(-\bar q(s))(1-\Gam)+\frac{\partial}{\partial s}(-\phibo(s))\bbw P]}{(1-\Gam)^2},
}
and similarly, 
\eq{
	&B_0\trieq\limso \frac{\frac{\Gam-\Gams}{1-\Gam}-\bar b(\jmath n\dome)}{\jmath n\dome}\\
= 	&\frac{\limso[(\frac{\partial}{\partial s}(-\overline{\Dphi}(s))(1-\Gamma)+\frac{\partial}{\partial s}(-\phibo(s))\frac{\Gam-\Gams}{1-\Gam}]}{(1-\Gam)^2}.
} 
These limits exist because, e.g., $\limso\frac{\partial}{\partial s}(-\phibo(s)) =\intoi t\phio(t)dt<\infty$ by A3. Thus, at frequency $0$, 
\eq{
	&\lim_{\ome\to0}\frac{\frac{\bbw P}{1-\Gam}-\bar a(\jw)}{\jw}\bar b(-\jw) = \frac{\Gam-\Gams}{1-\Gam}A_0,\\
	&\lim_{\ome\to0}\frac{\frac{\Gam-\Gams}{1-\Gam}-\bar b(\jw)}{\jw}\bar a(-\jw) = \frac{B_0}{1-\Gam}\bbw P.
	} 

{\bf Computation of $\ha(t)$: } 
Note that 
\eq{
\intoi a(t+u)b(t)dt=\wo{2\pi}\intii e^{u\jw}\bar a(\jw)\bar b(-\jw)d\ome.
} 
We have 
\eq{
\Wtil(u) &= \wo\pi\intii \cos(u\ome) \frac{\mathfrak{R}\{\bar q(\jw)\Dphib(-\jw)\}}{|1-\phibo(\jw)|^2}d\ome\\
&\approx \frac{\dome}{\pi}\ssum{n=-N_\ome}^{N_\ome}\cos(u\jmath n\dome)\frac{\mathfrak{R}\{\bar q(\jmath n\dome)\Dphib(-\jmath n\dome)\}}{|1-\phibo(\jmath n\dome)|^2},
}
where $\mathfrak R$ takes the real part. 
Generally, $\Wtil(u)$ is not analytical and, therefore, has to be 
estimated on a time grid on $[0,T]$ with grid width $\dt$. 
However, because of the $\cos$ component, 
at each time grid $m\dt$, we require that the frequency grid width $\dome\ll\frac{2\pi}{m\dt}$ to avoid Gibbs oscillation in the time domain. 
$\ha(t)$ is then computed via numerical convolution at each time grid over $[0,T]$.

{\bf Compution of the asymptotic covariances $\Sigoth,\Sigsth$: }
The previous computations do not require point process simulation. However, the asymptotic covariances involve the third moment. While a recursive expression for the third cumulant is available \cite{Jova15}, the resulting spectral integrals become a non-separable two-dimensional convolution, which is prohibitive to compute. We thus run Monte Carlo simulations to sample these expectations. 
We only need to sample $\Sigo=\E[\lamo(0)(\chi(0)-\mu)(\chi(0)-\mu)^\top]$ 
and $\Sigs = \E[\lamo(0)(\chi_h(0)-\muh)(\chi_h(0)-\muh)^\top]$. 
We simulate a large number $L$ of Hawkes process trajectories $N^l,l=1,\dotsm,L$ 
with a large observation time $T$ and sample 
\eq{
&\Sigo\approx \wo{L}\ssum{l=1}^L\lamtil_0(T;N^l)(\chitil(T;N^l)-\mu)(\chitil(T;N^l)-\mu)^\top\\
&\Sigs\approx \wo{L}\ssum{l=1}^L\lamtil_0(T;N^l)(\chitilh(T;N^l)-\mu)(\chitilh(T;N^l)-\mu)^\top,
} 
where $\lamtil_0(T;N^l)=c_0 + \intoTm \phio(T-u)dN^l_u$, 
$\chitil(T;N^l) = \intoTm q(T-u)dN^l_u$ and $\chitilh(T;N^l) = \intoTm q(T-u)+\ha(T-u)dN^l_u$. We can then compute $\Sigoth,\Sigsth$ using 
their formulas.

\end{document}